\documentclass[11pt,a4paper]{article}
\usepackage{epsfig,amsmath,amsfonts,verbatim}
\usepackage[top=0.7in,bottom=1in,left=1.1in,right=1.1in,includehead]{geometry}
\usepackage{lmodern}
\usepackage[T1]{fontenc}
\usepackage{marvosym,textcomp}
\usepackage{indentfirst}
\usepackage{amsfonts}
\usepackage{mathtools}
\usepackage[usenames,dvipsnames,svgnames,table]{xcolor}
\usepackage{tikz}
\usetikzlibrary{arrows}
\usetikzlibrary{shapes.misc}

\tikzset{cross/.style={cross out, draw=black, minimum size=2*(#1-\pgflinewidth), inner sep=0pt, outer sep=0pt},
cross/.default={5pt}}
 \usepackage[ normalem]{ulem}
 \usepackage{tocloft}
 
 \usepackage[debug,pageanchor=false]{hyperref}
\hypersetup{colorlinks=true,linktocpage,breaklinks,
            urlcolor=blue,
            linkcolor=red,
            citecolor=blue
            }

\numberwithin{equation}{section}

\def\a{\alpha} \def\b{\beta} \def\g{\gamma} \def\d{\delta} \def\e{\epsilon}
  \def\h{\eta} \def\q{\theta}
  \def\k{\kappa} \def\l{\lambda} \def\m{\mu}
\def\n{\nu} \def\x{\xi}   \def\r{\rho}
 \def\s{\sigma}   \def\f{\varphi}
\def\ff{\phi} \def\c{\chi} \def\y{\psi} \def\w{\omega}

\def\D{\Delta} 
   
   \def\L{\Lambda} 
 \def\X{\Xi}  
   
\def\F{\Phi}  \def\Y{\Psi} \def\W{\Omega}

\def\ba{\bar{a}}\def\bb{{\bar{b}}}\def\bc{\bar{c}}

\def\fr{\frac}  \def\dt{\partial}

\def\mc{\mathcal}

\def\XX{\mathbb{X}}

\def\TT{\mathbb{T}}

\newcommand\bqa {\begin{eqnarray}}
\newcommand\eqa {\end{eqnarray}}

\newcommand{\bear}{\begin{array}}
\newcommand{\enar}{\end{array}}

\newcommand{\un}[1]{\underline{#1}}

\def\beq{\begin{equation}}
\def\eeq{\end{equation}}
\def\bea{\begin{eqnarray}}
\def\eea{\end{eqnarray}}

\def\F{{\mathcal{F}}}

\def\GG{{\mathcal{G}}}
\def\V{\mathcal{V}}
\def\PP{\mathbb{P}}
\def\LL{\mathcal{L}}
\def\M{{\mathcal{M}}}
\def\DD{{\mathcal{D}}}

\begin{document}
\renewcommand{\contentsname}{}
\renewcommand{\refname}{\begin{center}References\end{center}}
\renewcommand{\abstractname}{\begin{center}\footnotesize{\bf Abstract}\end{center}} 
 \renewcommand{\cftdot}{}

\begin{titlepage}

\vfill

\begin{center}
   \baselineskip=16pt
      {\Large \bf  Exceptional field theory:  $SO(5,5)$}
      \vskip 2cm
       Aidar Abzalov${}^\star$\footnote{\tt arabzalov@edu.hse.ru}, Ilya Bakhmatov${}^\dagger$\footnote{\tt 
ivbahmatov@kpfu.ru
      }, Edvard T. Musaev${}^\star$\footnote{\tt emusaev@hse.ru}
          \vskip .6cm
                \begin{small}
                             {\it ${}^\star$National Research University Higher School of Economics,\\
                              Faculty of Mathematics \\
                             7, st. Vavilova, 117312, Moscow, Russia} \\[0.5cm]
                             {\it ${}^\dagger$Kazan Federal University, Institute of Physics \\
                             General Relativity Department\\
                              18, st. Kremlevskaya, 420008, Kazan, Russia} 
   \end{small}
\end{center}

\vfill 
\begin{center} 
\textbf{Abstract}
\end{center} 
\begin{quote}
We construct Exceptional Field Theory for the group $SO(5,5)$ based on the extended (6+16) -dimensional spacetime, 
which after reduction gives the maximal $D=6$ supergravity. We present both a true action and a duality-invariant 
pseudo-action formulations. All the fields of the theory depend on the complete extended spacetime. The U-duality group 
$SO(5,5)$ is made a geometric symmetry of the theory by virtue of introducing the generalised Lie derivative that 
incorporates a local duality transformation. Tensor hierarchy appears as a natural consequence of the algebra of 
generalised Lie derivatives that are viewed as gauge transformations. Upon truncating different subsets of the extra 
coordinates, maximal supergravities in $D=11$ and $D=10$ (type IIB) can be recovered from this theory.
\end{quote} 
\vfill
\setcounter{footnote}{0}
\end{titlepage}

\clearpage
\setcounter{page}{2}

\tableofcontents

\section{Introduction}

Recently the idea of a certain kind of geometry underlying the U-duality symmetries of toroidal 
compactifications of 11-dimensional supergravity~\cite{Cremmer:1978km} has gained a lot of attention. Since the seminal 
works~\cite{Cremmer:1978ds,Cremmer:1979up} it has been known that the field content of supergravities in lower dimensions can be organised into representations of the symmetry groups $E_{d}$ (for the $\mathbb{T}{}^d$ 
compactification) that appear to be the hidden symmetries of the theories. 

These symmetries have found their geometrical interpretation in the formalism of extended geometry, which has grown out from Hitchin's generalised geometry~\cite{Hitchin:2004ut,Gualtieri:2003dx} and its extension to exceptional symmetry groups~\cite{Hull:2007zu}. Building upon extended geometry techniques, development of double field theory~\cite{Siegel:1993th,Hull:2009mi,Hull:2009zb,Hohm:2010jy,Hohm:2010pp} and its extension to exceptional symmetry groups~\cite{Berman:2010is,Berman:2011cg,Berman:2011pe,Coimbra:2011ky,Coimbra:2012af} has brought forward the idea that not only the tangent space, but the target space itself becomes extended by introduction of a set of new coordinates $\XX{}^M$. From the point of view of string or M-theory these correspond to the winding modes of the extended objects, fundamental strings or M-branes. Essentially, construction of the extended geometry underlying the U-duality symmetry group $E_d$ of maximal supergravity compactified on a torus $\TT{}^d$ is based on the two simple principles:
\begin{itemize}
\item infinitesimal general coordinate transformations are replaced by generalised Lie derivatives that respect the $E_d$ structure;
\item the dynamics is restricted by a differential constraint called the section condition.
\end{itemize}
 The first principle may possibly allow one to consider non-geometric backgrounds, consistent from the point of view of 
string or M-theory, on the same footing as geometric 
ones~\cite{Dibitetto:2012ia,Dibitetto:2012rk,Hassler:2013wsa,Jensen:2011jna,Andriot:2012an}. The local dynamics of the 
theory is described by the so called generalised Lie derivative~\eqref{Lie}~\cite{Coimbra:2011ky,Berman:2012vc}, which 
combines the conventional translation term with and a $E_{d}$ transformation of a special form, plus a possible weight 
term. The section condition appears as a necessary constraint that must be included in order to keep the algebra of 
generalised Lie derivatives closed and to make it satisfy the Jacobi identity. This constraint is an extended geometry 
analogue of the level matching condition and its solutions correspond to different choices of the U-duality frame. 
Geometric structure of the extended space at finite distances is still not known in full detail, although there was 
certain progress in this direction \cite{Hohm:2012gk,Park:2013mpa,Lee:2013hma,Berman:2014jba,Cederwall:2014kxa}. 

This geometrical formalism appears as a basis for building the so-called Exceptional Field Theories (EFT), where the local duality transformations induced by the generalised Lie derivative act as gauge symmetries. These were constructed in the series of works \cite{Hohm:2013jma,Hohm:2013pua, Hohm:2013vpa,Hohm:2013uia,Hohm:2014fxa,Hohm:2015xna} for the groups $E_{6,7,8}$ and $SL(2)\times SL(3)$. The EFT's for the groups $E_{6,7}$ were further extended to include fermions in a supersymmetry invariant way in \cite{Godazgar:2014nqa,Musaev:2014lna,Musaev:2015pla}. Covariant gravitational field theory based on the $SL(N)$ extended space was constructed in \cite{Park:2013gaj,Blair:2013gqa,Park:2014una}. In this paper we continue building the chain and present the (bosonic) EFT for the group $SO(5,5)$ that corresponds to the maximal supergravity in $D=6$ spacetime dimensions.  In addition to the EFT generalisation of the true action constructed by Tanii~\cite{Tanii:1984zk} and used in \cite{Bergshoeff:2007ef} for maximal gauged supergravity, we construct a manifestly duality invariant pseudo-action and comment on their relationship. 

The full spacetime of a maximal $D$-dimensional supergravity is enlarged by inclusion of the extended space 
and all the fields now live on the full $(D+n)$-dimensional spacetime. Because of this natural 
split, the $D$ coordinates $x{}^\m$ are called external while the remaining $n$ coordinates $\XX{}^M$ 
are called internal. This is justified by the particular solution of the section constraint when 
the fields have no dependence on $\XX{}^M$, which corresponds to the reduction of 11-dimensional supergravity on a 
torus $\TT{}^{11-D}$. However, the structure of EFT is richer and we show that  it gives both 
11-dimensional supergravity and Type IIB supergravity as less trivial solutions of the section 
constraint.

The central pillar of EFT is the notion of covariant derivative along the external coordinates that 
respects the structure of extended geometry. Following the usual Yang-Mills like approach, the full 
content of the corresponding maximal supergravity becomes employed in the construction of covariant 
field strengths. Certain dual fields have to be added to the construction to ensure the covariance. 
We show that dynamics of the scalar sector, whose fields are encoded in the generalised metric, is 
determined by the so-called scalar potential, which is proven to be duality invariant, although 
written in a non-covariant form. Its truncation to the internal space was constructed in a series of 
works \cite{Berman:2011jh,Berman:2011kg,Berman:2011pe} and its geometrical meaning was investigated in \cite{Hohm:2011zr,Hohm:2011dv,Coimbra:2011ky,Coimbra:2011nw,Coimbra:2012af,Coimbra:2014uxa}.

It is worth mentioning that although the duality symmetries of supergravities were found in toroidal 
compactifications, the construction of extended geometry, and hence of EFT's, is not bound to this 
class of backgrounds. The torus is considered as a solution of equations of motion of EFT that 
preserves all duality symmetries and the full set of supersymmetries. One may be interested in 
searching for other solutions of EFT. Certain progress in this direction has been made in the works 
\cite{Cederwall:2014kxa} and \cite{Cederwall:2014opa}.

This paper is structured as follows. In the section \ref{Content} we describe the field content of the maximal $D=6$ supergravity, the dualisations necessary for the covariant construction and the pseudo-action formalism. In the section \ref{EG} the structure of extended geometry is briefly reviewed and basic algebraic identities needed further are provided.  In the section \ref{Der} we construct the covariant derivative and describe the tensor hierarchy in universal terms. The corresponding true action and the pseudo-action together with the Einstein-Hilbert term are presented in the section \ref{EFT}. Finally, in the section \ref{Embed} we consider the solutions of the section constraint that give the embedding of the 11-dimensional supergravity and Type IIB supergravity. Our conventions and notations and details of the most laborious calculations are collected in the Appendix.

\section{Field content and dualisations}
\label{Content}

The ungauged maximal 6-dimensional supergravity theory was originally constructed in~\cite{Tanii:1984zk}. Under 
the $6+5$ decomposition the metric and the 3-form of $D=11$ supergravity give rise to the following 
fields in the 6-dimensional theory ($m,n$ are internal indices running from 1 to 5):
\beq\label{split}
\left\{ g_{\m\n}, A_{\m\,m}, \ff_{mn}, C_{\m\n\r}, B_{\m\n\,m}, A_{\m\,mn}, \ff_{mnp}  \right\}.
\eeq
It is conventional to replace the 3-form $C_{\m\n\r}$ by the 1-form that is its dual in 6 
dimensions. Together with five 1-forms $A_{\m\,m}$ and ten 1-forms $A_{\m\,mn}$ this gives a total 
of sixteen 1-form fields, which are conveniently organized into a Majorana-Weyl spinor 
representation of the duality group $SO(5,5)$, $A_\m^M$, $M=1,\ldots,16$. The 2-form fields 
$B_{\m\n\,m}$ are in the $\mathbf{5}$ of $GL(5)\subset SO(5,5)$. Finally, the 25 scalar fields $\ff_{mn}, 
\ff_{mnp}$ are assembled into a 16 by 16 matrix $V_M{}^{\a\dot\a}$, which parameterises the coset 
$SO(5,5) / \left( SO(5)\times SO(5) \right)$, $\a,\dot \a = 1,\ldots,4$. 
This can be used to construct the generalised metric $\mc{M}_{MN}$ defined on the extended space:
\begin{equation}
\label{dM_SO}
\mc{M}_{MN} = V_M{}^{\a\dot\a}\, V_{N\,\a\dot\a},
\end{equation}
where the inverse scalar matrix is defined by
\begin{equation}
V_M{}^{\a\dot\a}\, V_{\a\dot\a}{}^N = \d_M{}^N,\qquad V_M{}^{\a\dot\a}\, V_{\b\dot\b}{}^M = \d_\b^\a\, 
\d_{\dot\b}^{\dot\a}.
\end{equation}
 The $SO(5)$ spinor indices are raised and lowered by the $\mathrm{USp}(4)$ 
invariant tensor $\W_{\a\b}$ which satisfies $\W_{\a\b}\W{}^{\b\g}=-\d{}^\g_\a$. This construction 
justifies calling the scalars $V_M{}^{\a\dot\a}$ the generalised vielbein.

In order to be able to account for the different possible gaugings of the $D=6$ theory, we 
introduce 
the duals of the 2-forms and the 1-forms as independent fields~\cite{Bergshoeff:2007ef}:
\begin{equation}
\left\{B_{\m\n}{}^m, C_{\m\n\r\, M}\right\}.
\end{equation}
From the point of view of the gauged theory, the additional five 2-forms $B_{\m\n}{}^m$ are added into the construction to incorporate the magnetic gaugings corresponding to the subgroups of the duality group $G$ which are not off-shell realised in the ungauged theory.  For theories in $D=4$ this was done in \cite{deWit:2005ub,deWit:2007mt}. Equations of motion for the magnetic 2-forms, which are considered independent, give Bianchi identities for the 3-form field strength, while the 3-form potentials give self-duality equations, restoring the correct amount of degrees of freedom. As it was shown in \cite{Bergshoeff:2007ef} in the six-dimensional theory this is possible only if gaugings are turned on. Alternatively, one may consider the exceptional field construction as it is done further.
 
While the Lagrangian itself is not duality invariant, the corresponding equations of motion can be recast into a duality covariant form. To this end, the magnetic and the electric 2-form potentials $B_m$ and $B{}^m$ are combined 
into the $\bf 10$ of $SO(5,5)$, which we denote by $B_{\m\n\,i}$, $i=1,\ldots,10$. In what 
follows it will be convenient to define
\begin{equation}
\begin{aligned}
B_{\m\n}{}^{KL} & = \fr{1}{16\sqrt{2}}\g{}^{i\, KL}B_{\m\n\, i},\\
C_{\m\n\r}{}^{M,KL}&=-\fr{1}{6\cdot 160}\g{}^{i\,KL}\g_i{}^{MN}C_{\m\n\r\, N}.
\end{aligned}
\end{equation}
The coefficients here were chosen so as to make the normalisation of the fields ${B}_{\m\n\, i}$ and ${C}_{\m\n\r\, M}$ the same as in~\cite{Bergshoeff:2007ef}.

This field content of the $SO(5,5)$ Exceptional Field Theory is in agreement with the analysis~\cite{Riccioni:2009xr} of decomposition of the $E_{11}$ representations under dimensional reduction. Under the $6+5$ decomposition we find the following representations $\mc{R}_p$ for $p$-forms:
\begin{equation}
\begin{aligned}
 \mc{R}_1 = \bf{16}, && \mc{R}_2=\bf{10}, && \mc{R}_3=\bf{\overline{16}}.
\end{aligned}
\end{equation}
The 4-forms are dual to scalars and do not appear as independent fields in the formalism. The 5-form potentials that live in the $\bf 144$ of $SO(5,5)$ are dual to mass deformations and are encoded in the embedding tensor, which naturally appears in the generalised Scherk-Schwarz reduction \cite{Berman:2012uy,Musaev:2013rq,Baron:2014yua}.

As we are working in even spacetime dimension $D=6$, we have to face a common subtlety when defining the action for the $(\frac{D}{2}-1)$-form potential and its dual.  Here one distinguishes between the genuine action and the so-called pseudo-action. The genuine action is not duality invariant itself, but the equations of motion may be cast into a duality covariant form by considering them on the same footing with Bianchi identities for the field strengths. Lagrangians of this kind were used by Tanii in his formulation of $D=6$ supergravity~\cite{Tanii:1984zk} as well as in~\cite{Bergshoeff:2007ef} in order to write down the gauged version of the theory. 

In its turn a pseudo-action is written completely in terms of fully $SO(5,5)$ covariant objects and is  invariant under the duality transformation. However, in order to compare the equations of motion one has to impose self-duality condition on the $SO(5,5)$ covariant 3-form field strength dressed up with scalar fields by hands.

Let us start with the kinetic term for the 2-form fields of the Tanii's action, which can be written in the following $GL(5)$ covariant form:
\begin{equation}
\label{true}
\mc{L}_T =- \fr{e}{2\cdot 3!} K^{mn} F^{\m\n\r}{}_m F_{\m\n\r n},
\end{equation}
where $e=\det e{}^{\ba}_\m$. This is a genuine action and it is written only for the field strengths of the five electric 2-forms $F_{\m\n\r\,m} = dB_{\m\n\, m}$. The matrix  $K^{mn}$ is built up from the scalar fields of the theory, and we are using the basis introduced in \cite{Bergshoeff:2007ef}:
\begin{equation}
K^{mn}=\V^{ma}(\V_n{}^a){}^{-1} P_+ - \V{}^{m\dot{a}}(\V_n{}^{\dot{a}}){}^{-1}P_-,
\end{equation}
where $P_{\pm}=1/2\,(1\pm *)$ is the projector on (anti)self-dual 3-forms and $*$ denotes the Hodge duality operator. Note, that one should understand the matrix $K^{mn}$ as an operator, acting only on 3-forms. The coset representative is written in the following $GL(5)\subset SO(5,5)$ covariant form:
\begin{equation}
V_M{}^{\a\dot\a}=
\begin{bmatrix}
\mc{V}_m{}^a & \mc{V}_m{}^{\dot{a}} \\
\mc{V}{}^{ma} & \mc{V}{}^{m{\dot{a}}} 
\end{bmatrix},
\end{equation}
where $a$ and $\dot{a}$ are the vector indices of $SO(5)\times SO(5)$. Such choice of the basis for the scalar matrix explicitly breaks $SO(5,5)$ covariance, preserving only its $GL(5)$ subgroup. This reflects the fact that the Lagrangian~\eqref{true} is not duality invariant.

Next, let us see how the equations of motion can be unified with the Bianchi identities in a duality covariant manner. To this end, one defines another 3-form $G_{\m\n\r}{}^m$ which is on-shell dual to the field strengths $F_{\m\n\r\, m}$ (see \cite{Gaillard:1981rj,Tanii:1998px} for reviews):
\begin{equation}
\label{duality}
*\!G^m =- \frac{3!}{e} \fr{\dt \mc{L}}{\dt F_m} = K^{mn} F_n;\qquad G^m = K^{mn} *\!F_n.
\end{equation}
Introducing a 10-plet of the 3-form field strengths as 
\begin{equation}
\label{10pletG}
G_{\m\n\r\,i}=
\begin{bmatrix}
F_{\m\n\r\,m}\\
G_{\m\n\r}{}^m
\end{bmatrix},
\end{equation}
the field equations and the Bianchi identities for $F_{\m\n\r\, m}$ can be written in an $SO(5,5)$ covariant form simply as $*dG_i=0$. We stress that the 3-form $G^m$ is defined by the equation \eqref{duality} and it is not considered as a field strength of some magnetic 2-form potential. However, the duality covariant equations of motion can be understood as coming from the following $SO(5,5)$-covariant variation
\begin{equation}
\d \tilde{\mc{L}}= dG_i \wedge \d B^i,
\end{equation}
where the variations $\d B_m$ and $\d B^m$ of the magnetic and electric 2-form potentials are considered as independent. This is precisely the idea behind the action  for $D=6$ maximal gauged supergravity and the formulation of the true action for $SO(5,5)$ exceptional field theory provided here.

To turn to the pseudo-action formulation it is convenient to represent the scalar matrix $K^{mn}$ as
\begin{equation}
K^{mn}=K_1^{mn}+K_2^{mn}*,
\end{equation}
where $K_1$ is symmetric and $K_2$ is antisymmetric. Then, the Lagrangian~\eqref{true} decomposes as
\begin{equation}
\mc{L}_T = -\fr{e}{2\cdot 3!} K_1^{mn} F_{\m\n\r\,m} F^{\m\n\r}{}_n-\fr{1}{2\cdot 3!3!} \e^{\m\n\r\s\k\l} K_2^{mn} F_{\m\n\r\,m} F_{\s\k\l\,n}.
\end{equation}
Consider now a 10-plet of 3-forms $F_i$ whose components $F_m$ and $F^m$ are completely independent on the level of the action and are understood as field strengths of the corresponding potentials
\begin{equation}
\label{10pletF}
F_{\m\n\r\,i}=
\begin{bmatrix}
F_{\m\n\r\,m}\\
F_{\m\n\r}{}^m
\end{bmatrix}.
\end{equation}
To be able to go back to five physical degrees of freedom  one introduces the  following self-duality relation by hands (for a more detailed discussion see \cite{Samtleben:2008pe} and \cite{Gaillard:1981rj}):
\begin{equation}
\label{self-d}
F_{\m\n\r\, i}=-\fr{1}{3!}\,e^{-1}\e_{\m\n\r\s\k\l}\,\h_{ij}\M^{jk}F^{\s\k\l}{}_k,
\end{equation}
where the symmetric matrix $\M_{ij}$ is built out of $K_1$ and $K_2$ as blocks in the following way:
\begin{equation}
\label{M_fix}
\mc M=-
\begin{bmatrix}
K_1 - K_2 K_1^{-1}K_2 & K_2 K_1^{-1}\\
-K_1^{-1}K_2 & K_1^{-1}
\end{bmatrix}.
\end{equation}
The $SO(5,5)$ invariant symmetric tensor $\h_{ij}$ is just a flat metric chosen to be
\begin{equation}
\h_{ij}=
\begin{bmatrix}
0 & \bf 1\\
\bf 1 & 0
\end{bmatrix}.
\end{equation}
The condition that the self-duality relation~\eqref{self-d} is invertible gives the following constraint for the scalar matrix:
\begin{equation}
\M_{ij}\h^{jk}\M_{kl}=\M_{il}.
\end{equation}

Now, the self-duality equation relates the magnetic components $F^m$ to the electric ones precisely in the same way as \eqref{duality}. Indeed, let us work in the matrix notation denoting $F_m$ and $F^m$ by $F_1$ and $F_2$ respectively. Then~\eqref{self-d} translates into
\begin{equation}
\begin{bmatrix}
F_1\\
F_2
\end{bmatrix}=
\begin{bmatrix}
0 & \bf{1} \\
\bf{1} & 0
\end{bmatrix}
\begin{bmatrix}
K_1 - K_2 K_1^{-1}K_2 & K_2 K_1^{-1}\\
-K_1^{-1}K_2 & K_1^{-1}
\end{bmatrix}
\begin{bmatrix}
*F_1 \\
*F_2
\end{bmatrix},
\end{equation}
that is
\begin{equation}
\begin{aligned}
F_1=&-K_1^{-1}K_2*\!F_1+K_1^{-1}*\!F_2,\\
F_2=&(K_1-K_2K_1^{-1}K_2)*\!F_1+K_2K_1^{-1}*\!F_2.
\end{aligned}
\end{equation}
Multiplying the first equation by $K_2$ from the left and subtracting the second one we obtain
\begin{equation}
\begin{aligned}
\label{dual}
F_2 = K_2 F_1 + K_1 *\!F_1, && \Longrightarrow && *F^m=K^{mn}F_n,
\end{aligned}
\end{equation}
where we used that the Hodge start squares to one acting on 3-forms in $D=6$ with Lorentzian signature, $*^2=+1$. Hence, under the self-duality condition \eqref{self-d} the magnetic 3-form field strength $F^m$ can be identified with the dual 3-form $G^m$.

Using the above relations one may show that the field equations of the genuine action~\eqref{true} together with the Bianchi identities  can be obtained by varying the following duality invariant pseudo-action:
\begin{equation}
\mc{L}_T=-\fr{1}{2\cdot 3!}\,\M^{ij}F_{\m\n\r\, i} F^{\m\n\r}{}_j,
\end{equation}
and imposing the constraint~\eqref{self-d}. Indeed, variation of the above action gives the following covariant equation of motion
\begin{equation}
*\!d *\! \M^{ij}F_j=0
\end{equation}
Imposing the self-duality constraint we obtain $*d G^i=0$, since the magnetic component $F^m$ becomes equal to the dual field strength $G^m$. With a more lengthy but straightforward calculation one can show that the above pseudo-action reproduces field equations for the scalar fields as well.

Note that the self-duality constraint has to be imposed after writing the field equation for pseudo-action. One may check that the pseudo-action itself as well as its variation vanish identically upon the self-duality condition.  Thus, the pseudo-action is not a reformulation of the true action but rather is a duality-invariant way to encode the equations of motion.

\section{Extended geometry}
\label{EG}

The transformation of tensors that is consistent with the structure of extended geometry is given by
\begin{equation}
\label{Lie}
\d_\L V{}^M = (\mc{L}_\L V){}^M = (L_\L V){}^M+Y{}^{MN}_{KL}\dt_N\L^K V{}^L\equiv[\L,V]_D^M, 
\end{equation}
where $[,]_D$ denotes the Dorfman bracket. Here both the transformation parameter $\L^M$ and the vector $V{}^M$ are functions of the extended coordinate $\XX{}^M$. Capital Latin indices run from 1 to $n$, which depends on the U-duality group under consideration. The tensor $Y{}^{MN}_{KL}$, which is an invariant tensor of the corresponding U-duality group, is essentially a projector \cite{Coimbra:2011ky}:
\begin{equation}
\label{Y}
\begin{array}{rcll}
O(d,d)_{strings}: & \quad & Y{}^{MN}_{KL}  = \eta{}^{MN} \eta_{KL}, & n=d, \vspace{0.2cm} \\
SL(5):  & \quad & Y{}^{MN}_{KL}= \e{}^{\a MN}\e_{\a KL}, & n=10, \\[0.2cm]
SO(5,5): &\quad & Y{}^{MN}_{KL}  = \frac{1}{2} (\g{}^i){}^{MN} (\g_i)_{KL} \ , &n=16, \\[0.2cm]
E_{6(6)}: &\quad & Y{}^{MN}_{KL}  = 10\, d{}^{MN R} d_{KLR} \ , &n=27, \\ [0.2cm]
E_{7(7)}: &\quad & Y{}^{MN}_{KL}  = 12\, c{}^{MN}{}_{KL} + \delta{}^{(M}_K \delta{}^{N)}_L + \frac{1}{2} \e{}^{MN} \e_{KL } \ &n=56. 
\end{array}
\end{equation}
Here the Greek indices $\a,\b,\g = 1,\ldots, 5$ label the representation $\bf{5}$ of $SL(5)$ and the index $i$ labels the $\bf{10}$ of $SO(5,5)$. \footnote{These notations are for this section only. For global notations see Appendix \ref{not}.} The invariant metric on $O(d,d)$ is denoted by $\h_{MN}$, $\e_{\a MN}=\e_{\a,\b\g,\d\e}$ is the $SL(5)$ alternating tensor, the matrices $\g{}^{i\,MN}$ are $16\times16$ off-diagonal blocks of the $SO(5,5)$ gamma-matrices in the Majorana-Weyl representation, and the tensors $d_{MNK}$ and $c{}^{MN}{}_{KL}$ are symmetric invariant tensors of $E_6$ and $E_7$ respectively.

The invariant tensor $Y_{KL}^{MN}$ is subject to several algebraic relations that ensure closure of the algebra \cite{Berman:2012vc}:
\begin{equation}
\begin{split}
\label{rel}
&Y{}^{(MN}_{KL}Y{}^{R)L}_{PQ}-Y{}^{(MN}_{PQ}\d{}^{R)}_{K}=0 \mbox{ , for $d\leq5$},\\
&Y{}^{MN}_{KL}=-\a_d\, P_{K}{}^{M}{}_{L}{}^{N}+\b_d\,\d{}^M_K\d{}^N_L+\d{}^M_L\d{}^N_K,\\
&Y{}^{MA}_{KB}\,Y{}^{BN}_{AL}=(2-\a_d)\,Y{}^{MN}_{KL}+(D\b_d+\a_d)\,\b_d\,\d{}^M_K\d{}^N_L+(\a_d-1)\,\d{}^M_L\d{}^N_K.
\end{split}
\end{equation}
Here $d=11-D$ is the number of compact dimensions and $P_A{}^B{}_C{}^D$ is the projector on the adjoint representation of the corresponding duality group. It is defined as $P_A{}^B{}_C{}^DP_D{}^C{}_K{}^L=P_A{}^B{}_K{}^L$ and $P_A{}^B{}_B{}^A=\mbox{dim}(\mathrm{adj})$. The coefficients $\a_d$ and $\b_d$ depend on the duality group and for the cases in question take numerical values 
$(\a_4,\b_4)=(3,\fr{1}{5})$, $(\a_5,\b_5)=(4,\fr{1}{4})$, $(\a_6,\b_6)=(6,\fr{1}{3})$. The last line in \eqref{rel} with $n=\d{}^A_A$ is a direct consequence of the second relation and the properties of the projector. In addition for the tensor $Y{}^{MN}_{KL}$ to be invariant the following identity must hold
\begin{equation}
\label{rel1}
Y{}^{KL}_{(PQ}\d{}^R_{N)}-Y{}^{KL}_{S(P}Y{}^{RS}_{QN)}=0.
\end{equation}
Using the expressions above it is useful to rewrite covariant derivative of a generalised vector in the following form
\begin{equation}
\d_\L V{}^M=(\mc{L}_\L V){}^M=\L^N\dt_N V{}^M-\a_d P{}^M{}_L{}^N{}_K\dt_N \L^K V{}^L + \b_d (\dt_K \L^K)V{}^M.
\end{equation}
Here the last term plays the role of a weight term, which could be added to any transformation. For a generalised vector that transforms as in \eqref{Lie} the weight is equal to $\b_d$. In general for a tensor with $k$ indices each transforming as in \eqref{Lie} the weight will be $k\b_d$. However, one may consider generalised tensors of any weight and, as we will see later, these are necessary for the EFT construction. 

 The second term in the expression above represents a projection of the term $\dt_N \L^K$ on the U-duality algebra, since in general it does not belong to the structure group $E_{d(d)}$. This in contrast to General Relativity where any non-degenerate matrix belongs to the structure group $GL(D)$ and one does not need a projector.

In addition one introduces a differential constraint on all fields in the theory that restricts the dependence on the extended coordinate $\XX{}^M$
\begin{equation}
Y^{MN}_{KL}\dt_M\otimes \dt_N=0.
\end{equation}
This extra condition in particular implies the existence of trivial generalised transformations given by $\L_0{}^M=Y{}^{MN}_{KL}\dt_N\c{}^{KL}$, for any $\c{}^{KL}$. Indeed, the generalised Lie derivative \eqref{Lie} of a vector field $V{}^M$ along the trivial vector field $\L_0{}^M$ reads
\begin{equation}
\d_{\L_0}V{}^M=Y{}^{NK}_{PQ}\Big(\dt_N\c{}^{PQ}\dt_KV{}^M+\fr12 \dt_{NK}\c{}^{PQ} V{}^M \Big)-\fr12 Y{}^{NP}_{KL}Y{}^{MK}_{RS}\dt_{NP}\c{}^{RS} V{}^L.
\end{equation}
It is straightforward to check that the parameter $\L_0{}^M$ itself transforms as a generalised vector. Closure of the algebra and the Jacobi identity hold up to a trivial transformation as well. Hence for the Jacobiator of generalised transformations we have
\begin{equation}
[\d_{\L_1},\d_{\L_2},\d_{\L_2}]=\d_{\L_0},
\end{equation}
where the RHS acts on any extended vector trivially up to section condition. For closure of the algebra we have
\begin{equation}
[\mc{L}_{\L_1},\mc{L}_{\L_2}]=\mc{L}_{[\L_1,\L_2]_E},
\end{equation}
which may be viewed as a definition of the E-bracket $[,]_E$. Explicitly this is given by
\begin{equation}
\label{brackets}
\begin{aligned}
[\L_1,\L_2]_E&=2\L_{[1}{}^N\dt_N\L_{2]}+Y{}^{MN}_{KL}\dt_N\L_{[1}{}^K\L_{2]}{}^L,\\
[\L_1,\L_2]_E&=[\L_1,\L_2]_D-\fr12Y{}^{MN}_{KL}\dt_N(\L_1{}^K\L_2{}^L).
\end{aligned}
\end{equation}
It is important to note that the E-bracket is antisymmetric while the Dorfman bracket is not. This will play a crucial role in the construction of tensor hierarchy starting from the covariant derivative to be defined in the next section. In what follows one finds important the following Jacobi identity for the E-bracket 
\begin{equation}
\label{Jac_E}
\left[\left[\L_{[1},\L_2\right]_E,\L_{3]}\right]_E{}^M=\fr16\, Y{}^{MN}_{KL}\dt_N \left( [\L_{[1},\L_2]_E{}^K\L_{3]}{}^L \right).
\end{equation}

\section{Covariant derivative for the D-bracket and tensor hierarchy}
\label{Der}

In the section~\ref{EG} we have presented the algebra of generalised Lie derivatives that closes on 
the E-bracket. In this construction the fields and the generalised diffeomorphism parameter $\L^M$ 
depend only on the extended coordinates $\XX{}^M$. We now regard these coordinates as internal in the 
spirit of Kaluza-Klein compactification. The fields and all the gauge parameters are now allowed to 
depend on the external spacetime coordinates, which we denote by $x{}^\m$. However, the corresponding 
derivative $\dt_\m$ is not a generalised scalar 
\begin{equation}
\d_\L\dt_\m V{}^M \neq \mc{L}_{\L}\left(\dt_\m V{}^M\right).
\end{equation}
In order to fix this we introduce a long spacetime derivative, covariant with respect to the D-bracket as in 
the ordinary Yang-Mills construction:
\begin{equation}
\mc{D}_\m=\dt_\m-\mc{L}_{A_\m}=\dt_\m - \left[A_\m,\bullet\ \right]_D,
\end{equation}
where the generalised vector field $A_\m^M$ plays the role of the gauge connection. We identify this 
gauge connection with the vector field of the corresponding maximal supergravity that always has 
exactly the desired number of degrees of freedom.

Covariance of the derivative $\mc{D}_\m$ with respect to the generalised Lie derivative implies 
the following transformation law of the gauge field $A_\m^M$:
\begin{equation}
\label{connection}
\d_\L A_\m^M=\dt_\m \L^M-[A_\m,\L]_D{}^M=\mc{D}_\m\L^M.
\end{equation}
Since D- and E-brackets differ by a trivial transformation (see \eqref{brackets})  the above choice 
is a matter of convention. Here we take the transformation in this form to keep the analogy with the 
conventional Yang-Mills construction.

As usual, the commutator of covariant derivatives defines the field strength of the gauge field:
\begin{equation}
\label{com1}
[\mc{D}_\m,\mc{D}_\n] =-\mc{L}_{\F_{\m\n}}, \qquad \F_{\m\n}{}^M =\ 2\,\dt_{[\m}A_{\n]}^M-[A_{\m},A_{\n}]_E{}^M-Y{}^{MN}_{KL}\dt_N B_{\m\n}{}^{KL}.
\end{equation}
Here the extra term with the field $B_{\m\n}$ was added  since the first two terms do not form a 
generalised vector under the gauge transformations. Note that this term does not contribute to the generalised Lie 
derivative in \eqref{com1} as it is a trivial transformation.

As in the maximal gauged supergravity, field strength for the 2-form potential $B_{\m\n}{}^{KL}$ is 
defined by  the Bianchi identity for the covariant field strength $\mc{F}_{\m\n}{}^M$:
\begin{equation}
\label{bianchi-2}
\begin{aligned}
3\,\mc{D}_{[\m}\mc{F}_{\n\r]}{}^M= &-Y{}^{MN}_{KL}\dt_N \F_{\m\n\r}{}^{KL},\\
\F_{\m\n\r}{}^{KL}  = &\ 3\,\mc{D}_{[\m}B_{\n\r]}{}^{KL}+
\fr{3}{D(1-2\b_d)}\,Y{}^{KL}_{PQ}\Big(A_{[\m}^{(P}\dt_\n
A{}^{Q)}_{\r]}-\fr13[A_{[\m},A_{\n}]_E{}^{(P}A_{\r]}{}^{Q)}\Big)\\
&-3\big(\dt_N C_{\m\n\r}{}^{N,KL} - Y{}^{KL}_{PQ}\,\dt_N C_{\m\n\r}{}^{Q,PN}\big),
\end{aligned}
\end{equation}
where again the terms in the last line were  added to make sure that the 3-form field strength is 
indeed covariant, i.e. $\d_\L \mc{F}_{\m\n\r}{}^{KL} = \mc{L}_\L \mc{F}_{\m\n\r}{}^{KL}$. This term will 
be constructed out of the next field in the tensor hierarchy, which is the 3-form 
$C_{\m\n\r}{}^{M,KL}$. As above, these terms do not contribute to the Bianchi identity since they 
vanish identically under the appropriate contraction with the $Y$ tensor.

Finally, we will find useful the Bianchi identity that gives the 4-form field strength: 
\begin{equation}
\label{bianchi3}
\begin{aligned}
4\,\mc{D}_{[\m}\F_{\n\r\s]}{}^{KL}=&\fr{3}{D(1-2\b_d)}Y{}^{KL}_{PQ}\,\F_{[\m\n}{}^P\F_{\r\s]}{}
{}^Q-3\left(\dt_N
\F_{\m\n\r\s}{}^{N,KL}-Y{}^{KL}_{PQ}\,\dt_N \F_{\m\n\r\s}{}^{Q,PN}\right).
\end{aligned}
\end{equation} 
Substituting the explicit form of the fields into this expression we obtain the 4-form:
\begin{equation}
\begin{aligned}
\F_{\m\n\r\s}{}^{M,KL}=&\ 4\,\mc{D}_{[\m}C_{\n\r\s]}{}^{M,KL}+\left(2B_{[\m\n}{}^{KL}\F_{\r\s]}{}^{M}- B_{[\m\n}{}^{KL}Y{}^{MN}_{PQ}\dt_NB_{\r\s]}{}^{PQ}\right)\\
&+\fr{4}{3D(1-2\b_d)} 
Y{}^{KL}_{PQ}\left(A_{[\m}^M A_\n^P \dt_\r A_{\s]}^Q - \fr14 A_{[\m}^M [A_\n,A_\r]_E{}^P A_{\s]}^Q\right).
\end{aligned}
\end{equation}
Here one does not need to add any extra fields to covariantise the expression since it does not appear in the Lagrangian. Moreover, all possible extra terms should disappear from the Bianchi identity as well and hence do not show up at all. Due to the duality relation between $\F_{\m\n}$ and $\F_{\m\n\r\s}$ to be derived later as the field equation of the magnetic 2-form potential $B_{\m\n}{}^m$, one can also write down the external diffeomorphisms for the 3-form potential $C_{\m\n\r}$ using the 2-form field strength, rather than the 4-form.

Under arbitrary variations of the $p$-form potentials the covariant field strengths transform as 
follows:
\begin{equation}
\label{varF}
\begin{aligned}
\d \F_{\m\n}{}^M=&\ 2\,\mc{D}_{[\m}\D A_{\n]}^M-Y{}^{MN}_{KL}\dt_N \D B_{\m\n}{}^{KL},\\
\d \F_{\m\n\r}{}^{KL}=&\ 3\,\mc{D}_{[\m}\D B_{\n\r]}{}^{KL}+\fr{3}{D(1-2\b_d)}Y{}^{KL}_{PQ}\,\F_{[\m\n}{}^P\D 
A_{\r]}^Q \\
&-3\big( \dt_N \D C_{\m\n\r}{}^{N,KL} - Y{}^{KL}_{PQ}\, \dt_N \D C_{\m\n\r}{}^{Q,PN}\big),\\
\d \F_{\m\n\r\s}{}^{M,KL}=&\ 4\, D_{[\m}\D C_{\n\r\s]}{}^{M,KL}+\fr{1}{3 D(1-2\b_d)} \Big(\fr{3}{8}\F_{[\m\n}{}^{M}\D B_{\r\s]}{}^{KL}-\fr{1}{4}\F_{[\m\n\r}{}^{KL}\d A_{\s ]}^M\Big)
\end{aligned}
\end{equation}
where it proves useful to define ``covariant'' transformations 
\begin{equation}
\begin{aligned}
\D A_\m^M=&\ \d A_\m^M,\\
\D B_{\m\n}{}^{KL}=&\ \d B_{\m\n}{}^{KL}-\fr{1}{D(1-2\b_d)}Y{}^{KL}_{MN}A_{[\m}^{M}\d A_{\n]}^{N},\\
\D C_{\m\n\r}{}^{N,KL} =&\  \d C_{\m\n\r}{}^{N,KL} - \d A_{[\m}^N B_{\n\r]}{}^{KL} - \fr{1}{3D(1-2\b_d)}
Y{}^{KL}_{RS} A_{[\m}^N A_\n^R \d A_{\r]}^S.
\end{aligned}
\end{equation}
Identifying the field $B_{\m\n}{}^{KL}$ with the 2-form $B$-field of the maximal $D=6$ 
supergravity, we may expect its own gauge variation with a 1-form parameter $\X_\m{}^{KL}$ to 
appear in the transformation law as $\D B_{\m\n}{}^{KL}=2\mc{D}_{[\m}\X_{\n]}{}^{KL}+\mbox{other 
terms}$. This will make the variations~\eqref{varF} covariant. Apparently, the gauge variation of $A_\m^M$ would also be affected, and the same is true 
for the 3-form potential. Hence, requiring that the field strengths transform covariantly leads to the following gauge 
transformations of the fields corresponding to the $SO(5,5)$ duality 
group:\footnote{It is important to note a subtlety that arises in even dimensions. For the 
off-shell formulation of the theory the field $\F_{\m\n\r}{}^{KL}$ in the last line of \eqref{trans_AB} 
should be replaced by $\GG_{\m\n\r}{}^{KL}$. }
\begin{equation}
\label{trans_AB}
\begin{aligned}
\D A_\m^M =&\ \mc{D}_\m\L^M+Y{}^{MN}_{KL}\dt_N\X_\m{}^{KL},\\
\D B_{\m\n}{}^{KL} =& \
2\mc{D}_{[\m}\X_{\n]}{}^{KL}-\fr{1}{D(1-2\b_d)}Y{}^{KL}_{MN}\L^M\mc{F}_{\m\n}{}^N\\
&+3\left(\dt_N\Y_{\m\n}{}^{N,KL}-Y{}^{KL}_{PQ}\dt_N \Y_{\m\n}{}^{P,NQ}\right),\\
\D C_{\m\n\r}{}^{M,KL}=&\ 3\mc{D}_{[\m} 
\Y_{\n\r]}{}^{M,KL}-\mc{F}_{[\m\n}{}^M\X_{\r]}{}^{KL}+\fr{2}{3D(1-2\b_d)}Y{}^{KL}_{PQ}\L^P\mc{F}_{\m\n\r}{}^
{
QM}.
\end{aligned}
\end{equation}
In what follows we explicitly determine the relation between the field  $C_{\m\n\r}{}^{N,KL}$ in the 
formalism above and the  3-form potentials of the corresponding gauged supergravities. These have  
different structures of the indices and will be related by the $SO(5,5)$ invariant tensors.

In order to compare with \cite{Bergshoeff:2007ef}, one has to use the identity for $SO(5,5)$ gamma-matrices 
\begin{equation}
\label{gamma_id}
\g{}^{i(MN}\g_i{}^{K)L}=0.
\end{equation}
Then it is possible to rewrite the above relations in terms of the fields $B_{\m\n\, i}$ and $C_{\m\n\r\, M}$:
\begin{equation}
\label{trans_AB0}
\begin{aligned}
\D A_\m^M & = \mc{D}_\m \L^M + \fr{1}{2\sqrt{2}} \, \g{}^{i\,MN} \dt_N \X_{\m\, i},\\
\D B_{\m\n\, i} &= 
2\,\mc{D}_{[\m}\X_{\n] i}-\sqrt{2}\,\g_{i\,MN}\L^M\mc{F}_{\m\n}{}^N-\fr{\sqrt{2}}{4}\,\g{}^{i\,MN}\dt_M \Y_{\m\n\, N},\\
\D C_{\m\n\r\, M}&=3\,\mc{D}_{[\m}
\Y_{\n\r] M}+3\sqrt{2}\,\g{}^{i}{}_{MN}\mc{F}_{[\m\n}{}^N\X_{\r]i}+\sqrt{2}\,\g{}^i{}_{MN}\L^N\mc{F}_{\m\n\r\, i}.
\end{aligned}
\end{equation}
For the Bianchi identities we obtain:
\begin{equation} 
\label{bianchi}
\begin{aligned}
3\,\DD_{[\m}\F_{\n\r]}{}^M&=-\fr{1}{2\sqrt{2}}\,\g{}^{i\,MN}\dt_M \F_{\m\n\r\, i},\\
4\,\DD_{[\m} \F_{\n\r\s] i}&=3\sqrt{2}\,\F_{[\m\n}{}^{M}\F_{\r\s]}{}^N\g_{i\,MN}+\fr{\sqrt{2}}{4}\,\g_i{}^{MN}\dt_M \F_{\m\n\r\s\, N}.
\end{aligned}
\end{equation}

The covariant gauge transformation $\d_\L \F_{\m\n\r}{}^{KL}$ implies that the 3-form field strength is a rank 2 generalised tensor of weight $\l(\F_{(3)})=1/2$. Indeed, decomposing the $Y$-tensor in terms of the projector one obtains
\begin{equation}
\d_\L \F_{\m\n\r}{}^{KL}=\L^N\dt_N \F_{\m\n\r}{}^{KL}-8\, \F_{\m\n\r}{}^{Q(L} \, \PP{}^{K)}{}_Q{}^N{}_P \, \dt_N \L^P +\fr12\, \dt_N \L^N \F_{\m\n\r}{}^{KL}.
\end{equation} 
In what follows we will need the gauge transformation of the corresponding ${\bf10}$-plet $\F_{\m\n\r\, i}$, which takes the following suggestive form:
\begin{equation}
\label{10plet}
\d_\L \F_{\m\n\r\, i} = \L^N\dt_N \F_{\m\n\r\, i}-\fr12(t_i{}^j)_M{}^N \,\dt_N \L^M \F_{\m\n\r\, j} +\fr12\, \dt_N \L^N \F_{\m\n\r \,i},
\end{equation}
where $(t_{ij})_M{}^N=\g_{[i\,MP}\g_{j]}{}^{PN}$ represents the generators of $SO(5,5)$ in terms of the gamma-matrices. Here we have used the following identity
\begin{equation}
\PP{}^K{}_L{}^P{}_Q \, \g{}^{i\,QR}\g_{j\,RP}=(t{}^i{}_j){}^K{}_L,
\end{equation}
which is true since the left hand side is traceless with respect to 10-dimensional indices. Note that the expression \eqref{10plet} again has the form of a translational term plus weight plus an $SO(5,5)$ local duality rotation.

\section{Covariant exceptional field theory}
\label{EFT}

In this section we present the invariant Lagrangian for the $SO(5,5)$ Exceptional Field Theory, 
which has the following  schematic structure:
\begin{equation}\label{action}
\begin{aligned}
\LL_{EFT}=&\ \LL_{EH}(\hat R)+\LL_{sc}(\mc{D}_\m 
\M_{MN})+\LL_{V}(\F_{\m\n}{}^M)+\LL_{T}(\F_{\m\n\r}{}^{KL})\\
&+\LL_{top}-eV(\M_{MN},g_{\m\n}).
\end{aligned}
\end{equation}
Here the Einstein-Hilbert term $\LL_{EH}$, the kinetic term for the scalar fields $\LL_{sc}$ and the 
vector fields potential $\LL_{V}$ can be written in a duality covariant form. In contrast, the kinetic 
term for the rank 2 tensor potential $\LL_{T}$ as well as the topological Lagrangian  $\LL_{top}$ should be 
considered on a separate basis. Due to the usual subtlety with $(k-1)$-forms in even $D=2k$ dimensions, writing the Lagrangian for the 2-form potential in $D=6$ in a fully duality covariant manner is nontrivial. This can be achieved by giving up manifest Lorentz invariance~\cite{Henneaux:1988gg}, or by introducing extra scalar fields~\cite{Pasti:1996vs,DePol:2000re}. However, for our needs only the variation of the corresponding kinetic and topological Lagrangians is enough. As will be shown here, the extended geomtery allows to write this variation in a duality and Lorentz covariant way.

Finally, one should include the potential term $eV(\mc{M}_{MN},g_{\m\n})$ for the scalar fields, which
depends on derivatives along $\XX{}^M$ and transforms as a density under the generalised Lie derivative, 
leaving the action invariant.

\subsection{Universal kinetic Lagrangian}

For the curvature of the external metric $R_{\m\n\r\s}$ to be a scalar of weight zero under the 
gauge transformations induced by the generalised Lie derivative, the corresponding spin-connection 
$\w_\m{}^{\ba\bb}$ should have weight zero as well.  To ensure this we set the external vielbein to be a 
scalar of weight $\l(e{}^{\ba}_\m)=\b_d$. The usual equation that determines the spin-connection can be 
written in the following covariant form:
\begin{equation}
\mc{D}_{[\m} e_{\n]}{}^{\ba}-\fr14 \w_{[\m}{}^{ab}e_{\n]b}=0.
\end{equation}
In addition, since all the fields are dependent on the extended coordinates, so are the parameters 
$\L^a{}_b$ of Lorentz rotations. The corresponding Lorentz-invariant Riemann scalar then differs 
from the usual expression and has the same form as in \cite{Hohm:2013vpa}:
\begin{equation}
\hat{R}_{\m\n \ba\bb}=R_{\m\n \ba\bb}+\F_{\m\n}{}^M \, e{}^\r_{\ba}\, \dt_M e_{\r\, \bb}.
\end{equation}
Hence, the full covariant Einstein-Hilbert term takes the following form:
\begin{equation}
S_{EH}=-\fr12\int d{}^n x\, d{}^D\XX\, e \hat{R}= -\fr12\int d{}^n x \, d{}^D\XX \, e \, e{}^\m_{\ba} e{}^\n_{\bb}
\hat{R}_{\m\n}{}^{\ba\bb}.
\end{equation}

For the scalar degrees of freedom parameterised by the matrix $\mc{M}_{MN}$ one writes the general 
form of the Lagrangian as
\begin{equation}
\LL_{sc}=\fr{1}{4\a_d}\, e\,g{}^{\m\n} \, \mc{D}_\m \M_{MN} \, \mc{D}_\n \M{}^{MN}.
\end{equation}
This expression is explicitly covariant with respect to the local gauge transformation generated by the 
generalised Lie derivative. Since we have for the weight of the vielbein $\l(e{}^{\ba}_\m)=\b_d$, the 
total weight counting gives $(d-2)\b_d=1$, which is in precise correspondence with the pattern for 
$\b_d$ noticed in \cite{Berman:2012vc}. Indeed, if an expression $T$ has weight $\l(T)=1$, then its 
transformation can be written as a full derivative:
\begin{equation}
\d_\L T=\L^N \dt_N T + \l(T)\, \dt_N \L^N T=\dt_N(\L^N T).
\end{equation}
This will prove useful in the verification of gauge invariance of the potential term \\
$eV(\mc{M}_{MN},g_{\m\n})$.

The kinetic term for the 1-form potential $A_\m^M$ takes the following universal form:
\begin{equation}
\LL_{V}=-\fr14\, e\, \M_{MN}\, \F_{\m\n}{}^M \F_{\m\n}{}^N.
\end{equation}
One can substitute \eqref{dM_SO} for the scalar 
matrix $\M_{MN}$. Again, counting of weights gives the total weight of 1.

Hence, altogether we have for the kinetic terms that can be written in a universal form: 
\begin{equation}
\LL{}^{(U)}_{kin}=-\fr12 \,e \, \hat{R}[g,\F] + \fr{1}{4\a_d} \,e \, g{}^{\m\n}\, \mc{D}_\m \M_{MN} \,\mc{D}_\n \M{}^{MN} -\fr14\, e\, \M_{MN}\, \F_{\m\n}{}^M \F_{\m\n}{}^N.
\end{equation}
Because of the dualisation in even dimensions one has to consider the kinetic term for the 2-form potential separately. This term together with the corresponding topological Lagrangian is considered in the next two sections.

\subsection{Kinetic and topological action for the p-forms}

Comparing the transformation of the 2-form field \eqref{trans_Fmn} with that of 
\cite{Bergshoeff:2007ef} we define the following fields in the $\bf 10$ and $\bf \bar{5}$ 
representations:
\begin{equation}
\begin{aligned}
B_{\m\n}{}^{KL} & = \fr{1}{16\sqrt{2}}\g{}^{i\, KL}B_{\m\n i},\\
C_{\m\n\r}{}^{M,KL}&=-\fr{1}{6\cdot 160}\g{}^{i\,KL}\g_i{}^{MN}C_{\m\n\r\, N}.
\end{aligned}
\end{equation}

In analogy with the prescription of the gauged maximal $D=6$ supergavity we do the following 
replacements:
\begin{equation}
\begin{aligned}
F_{\m\n\r\, m} &\to \F_{\m\n\r\, m},\\
F_{\m\n}{}^M &\to \F_{\m\n}{}^M.
\end{aligned}
\end{equation}
It is important to note that the replacement $F_{\m\n\r\, m} \to \F_{\m\n\r\, m}$ only refers to the 5 of the 10 
components of the field $\F_{\m\n\r\, i}$. The remaining dual components will be restricted by the field equation of the 3-form field $C_{\m\n\r\,M}$. Hence, as was described in Section \ref{Content} the covariant on-shell 10-plet field strength  becomes
\begin{equation}
\mc{G}_{\m\n\r\,i}=
\begin{bmatrix}
\mc{G}_m \\
\mc{G}{}^m
\end{bmatrix}_{\m\n\r}
=
\begin{bmatrix}
\F_m \\
*K{}^{mn}\F_n
\end{bmatrix}_{\m\n\r}.
\end{equation}
 
Now we are able to write the full variation of the kinetic and topological Lagrangians for the $p$-forms with respect to variations of the $p$-form potentials \eqref{trans_AB} as follows
\begin{equation}
\label{delta_L1}
\begin{aligned}
&\d(\mc{L}_{\rm kin}+\mc{L}_{\rm top})=\\
=&-\fr e2\, \mc{M}_{MN}\F{}^{\m\n\, M}\d 
\F_{\m\n}{}^N - \fr{\k}{3!}\, \e{}^{\m\n\r\s\k\l}\, \h{}^{ij}\mc{G}_{\m\n\r\, i}\, \mc{D}_\s \D B_{\k\l\, j}\\
&-\fr{\sqrt{2}\,\k}{3!}\, \e{}^{\m\n\r\s\k\l}\,\mc{G}_{\m\n\r\, i}\, \g{}^{i}{}_{MN}\F_{\s\k}{}^M\d 
A_\l{}^N + \fr{\sqrt{2}\,\k}{8}\, \e{}^{\m\n\r\s\k\l}\, \F_{\m\n}{}^M\,\g{}^i{}_{MN}\,\F_{\r\s}{}^N\,\D B_{\k\l 
i}\\
&+\fr{\sqrt{2\,\k}}{3\cdot 4!}\, \e{}^{\m\n\r\s\k\l}\,(\F_{\m\n\r\, i}-\GG_{\m\n\r\, i})\,\g{}^{i\,MN}\dt_M \D C_{\s\k\l\, N}.
\end{aligned}
\end{equation}
Although we are working with the true action that is not duality invariant, this variation gives duality covariant equations of motion for the $p$-form field potentials. Note, that variations of the magnetic and electric 2-form potentials are considered to be independent, while the field strength $\GG_{\m\n\r\, i}$ contains only electric degrees of freedom. This is done to obtain the duality-covariant equations of motion with the correct number of physical fields. The magnetic degrees of freedom are encoded in the field strength $\F_i$ defined as
\begin{equation}
\F_{\m\n\r\, i}=
\begin{bmatrix}
\F_{\m\n\r\, m} \\
\F_{\m\n\r}{}^m.
\end{bmatrix}
\end{equation}
The duality relation restricting $\F^m$ will follow from the equations of motion of the 3-form potential $C_{\m\n\r\,M}$.

The above variation is constructed in the following way. One starts with the first two terms above 
with an arbitrary relative coefficient $\k$. These simply correspond to variations coming from the kinetic 
terms for the 1- and 2-form potentials. Next, one adds the necessary contributions to make the expression invariant under the gauge transformations generated by $\X_{\m\, i}$ and $\Y_{\m\n\, M}$. The most straightforward way to see this invariance is to rewrite the above expression using the equation \eqref{var_cov_C} as follows:
\begin{equation}
\label{delta_L1_simp}
\begin{aligned}
&\d(\mc{L}_{\rm kin}+\mc{L}_{\rm top})=\\
=&- \fr e2\, \mc{M}_{MN}\,\F{}^{\m\n\, M}\,\d 
\F_{\m\n}{}^N - \fr{\k}{3 \cdot 3!}\, \e{}^{\m\n\r\s\k\l}\, \h{}^{ij}\GG_{\m\n\r\, i}\, \d \F_{\s\k\l\, 
i}\\
&+\fr{\sqrt{2}\,\k}{8}\, \e{}^{\m\n\r\s\k\l}\, \F_{\m\n}{}^M\, \g{}^i{}_{MN}\,\F_{\r\s}{}^N\,\D B_{\k\l\,
i}-\fr{\sqrt{2}\,\k}{3\cdot 4!}\, \e{}^{\m\n\r\s\k\l}\,\F_{\m\n\r\, i}\,\g{}^{i\,MN}\dt_M \D C_{\m\n\r\, N},
\end{aligned}
\end{equation}
The first two terms are trivially invariant under the variations $\X_{\m\n\,i}$ and $\Y_{\m\n\r\,M}$ 
of the 2- and 3-forms respectively. To see that the $\X_{\m\n\, i}$ variations of the other two  
terms 
cancel, one integrates by parts $\dt_M$ in the second term and uses the Bianchi identity 
\eqref{bianchi2}. This gives a full derivative of the form $\DD(\F\F\,\X)$ and hence vanishes. 
Cancellation of $\Y_{\m\n\r\, M}$ variations works in the very same way. Note that $\X_{\m\n\, i}$ is a 
generalised $\bf 10$-plet of weight $\l_\X=1/2$ (cf. \eqref{10plet}).

Let us look at the equations of motion for the  3-form potential 
$C_{\m\n\r\, M}$ which give a relation between the 
covariant  field strength $\F_{\m\n\r\, i}$ and $\GG_{\m\n\r\, i}$:
\begin{equation}
\label{FmG}
\g_m{}^{MN}\,\dt_N(\F_{\m\n\r}{}^m-\GG_{\m\n\r}{}^m)=0.
\end{equation}
This is the EFT analogue of the equation
\begin{equation}
g\,\q{}^M_m(\F{}^m-*K{}^{mn}\F_n)=0,
\end{equation}
which constrains the dual component $\F_{\m\n\r}{}^m$. The above equation can be obtained from its EFT 
analogue by means of Scherk-Schwarz reduction, which expresses the components of the embedding tensor 
$\q{}^{M\,i}=(\q{}^{M\,m},\q{}^M_m)$ in terms of twist matrices. Covariance of the equation \eqref{FmG} in 
the extended geometry sense follows from the identity
\begin{equation}
\d_\L\Big(\dt_N \Y{}^N{}_{QR} -Y{}^{KL}_{P(Q}\dt_{R)}\Y{}^P{}_{KL}\Big)=\mc{L}_\L\Big(\dt_N \Y{}^N{}_{QR} 
-Y{}^{KL}_{P(Q}\dt_{R)}\Y{}^P{}_{KL}\Big),
\end{equation}
which is true for any generalised tensor $\Y{}^P{}_{KL}=\Y{}^P{}_{LK}$. 

Using the Bianchi identity \eqref{bianchi3}, the bosonic field equation of the magnetic 2-form 
potential $B_{\m\n}{}^m$ can be written in the following form
\begin{equation}
\label{dual_C}
\g_{m}{}^{KL}\, \dt_K\Big(\F_{\m\n\r\s\, L}+\fr{1}{4\k}\, \e_{\m\n\r\s\k\l}\, e\, \M_{LN}\,\F{}^{\k\l\, 
N}\Big)=0.
\end{equation}
This is the EFT analogue of the on-shell duality relation between the 3-forms and the 1-forms  (see the Section~\ref{Content}). This equation will prove useful for establishing invariance of the Lagrangian under 
external $d=5+1$ diffeomorphisms, that will fix all the remaining freedom in choosing relative 
coefficients in \eqref{delta_L1}. 

The relative factor $\k$ can not be fixed by gauge invariance and remains undetermined here. 
Further we will see that in order to have the Lagrangian invariant under the external $(5+1)$-dimensional 
diffeomorphisms generated by the shift $x{}^\m\to x{}^\m+\x{}^\m(x)$, one should set $\k=1/2$.

\subsection{Field equations and pseudo-action}

In the previous section the true action has been constructed. In its general form it repeats the action of maximal $D=6$ gauged supergravity, however with additional subtleties due to dependency on the extended coordinates. However, in order to provide a fully duality-covariant formulation of the theory one has to construct a pseudo-action.

The kinetic term for 1- and 2-form potentials has its usual form and can be easily written as
\begin{equation}
\mc{L}_{kin}=-\fr{e}{2\cdot 3!}\F_{\m\n\r\, i}\M^{ij}\F^{\m\n\r}{}_j-\fr{e}{4}\F_{\m\n}{}^M\F^{\m\n\, N}\M_{MN},
\end{equation}
where $\M^{ij}$ is the $10\times 10$ duality covariant scalar matrix constructed of the matrices $K_1^{mn}$ and $K_2^{mn}$ as blocks (see Section~\ref{Content} and the lectures \cite{Samtleben:2008pe} for more details). Here we have already set $\k=1/2$ for convenience. In addition, to obtain equations of motion consistent with the first order self-duality equations and Bianchi identities one should add a topological term, that is a term that does not contain the spacetime metric $g_{\m\n}$ as well as the scalar matrices $\M_{MN}$ or $\M^{ij}$. As in the gauged case the easiest way to do this is to construct its variation, since the topological Lagrangian itself is not covariant. Hence, we have
\begin{equation}
\label{var_top}
\begin{aligned}
\d \mc{L}_{top}=&\ e^{-1}\e^{\m\n\r\s\k\l}\Big(-\fr{1}{3!}\DD_\s \F_{\m\n\r i}\D B_{\k\l}{}^i+\fr{1}{3\sqrt{2}}\F_{\m\n\r i}\F_{\s\k}\g^i\d A_\l\\
&-\fr{1}{8\sqrt{2}}\F_{\m\n}\g^i\F_{\r\s}\D B_{\k\l\, i}+\fr{1}{4!3\sqrt{2}}\F_{\m\n\r\, i}\g^{iMN}\dt_M \D C_{\m\n\r\, N}\Big),\\
\end{aligned}\end{equation}
where we have used the spinor notation for $\F_{\m\n}{}^M$ and $\d A_\m^M$. Given the expressions \eqref{trans_AB0} it is straightforward to show that the above variation vanishes on the gauge transformation. Hence, the corresponding pseudo-action is duality invariant. Note, that this topological term has very similar structure to the one obtained in \cite{Samtleben:2011fj, Bandos:2013jva}.

Now using the general variations of the field strengths \eqref{varF} the above variation can be recast in the following nice covariant form
\begin{equation}
\begin{aligned}
\d\mc{L}_{top}&= e^{-1}\e^{\m\n\r\s\k\l}\Big(\fr{1}{36}\F_{\m\n\r\, i}\,\d \F_{\s\k\l}{}^i+\fr{1}{48}\F_{\m\n\r\s\, M}\,\d \F_{\k\l}{}^M\Big)\\
&=\F^i\wedge\d\F_i+\F_M\wedge \d \F^M,
\end{aligned}
\end{equation}
where we define a $p$-form $\w$ as
\begin{equation}
\w=\fr{1}{p!}\w_{\m_1\ldots \m_p}dx^{\m_1} \wedge \cdots \wedge dx^{\m_p}.
\end{equation}
Using the explicit form of the variations \eqref{varF} together with Bianchi identities \eqref{bianchi} after a lengthy but straightforward calculation one shows that the variation $\d (\mc{L}_{kin}+\mc{L}_{top})$ gives the same equations of motion as the true action \eqref{delta_L1} upon the self-duality condition that is imposed by hands.

 It is a common situation for  Exceptional Field Theories that the topological term is most conveniently written as an integral of a full derivative over a higher-dimensional space whose boundary is the 6-dimensional spacetime.\footnote{Note, that this is just a convenient way to encode the topological term and to reproduce its variation. There is no physical meaning of the $D=7$ spacetime in this setting.} With some abuse of notation this can be written as
\begin{equation}
\label{L_top}
\begin{aligned}
S_{top}&=\int d^6x\, d^{16}\XX\, \mc{L}_{top}\\
&=\int d^7X\, d^{16}\XX\, \left(2\,\h^{ij}\F_i\wedge \DD\F_j-\fr{1}{\sqrt{2}}\F\wedge\g^i\F\wedge \F_i\right)
\end{aligned}
\end{equation}
where we used the following differential form notation
\begin{equation}
\begin{aligned}
\F^M&=\fr{1}{2}\F_{\m\n}{}^M dX^\m \wedge dX^\n ,\\
\F_i&=\fr{1}{3!}\F_{\m\n\r\, i}\, dX^\m \wedge dX^\n \wedge dX^\r .
\end{aligned}
\end{equation}
Again the above expression is very similar to the structure of the topological action of \cite{Samtleben:2011fj, Bandos:2013jva}.

The particular form of the topological Lagrangian $\mc{L}_{top}$ is not manifestly covariant and therefore is not very useful for our further discussion. Invariance of the topological action as well as equivalence of the variation of \eqref{L_top} to \eqref{var_top} goes precisely in the same way as for the $E_{7(7)}$ and $SL(2)\times SL(3)$ exceptional field theories \cite{Hohm:2013uia,Hohm:2015xna}. Note that each term in the topological action \eqref{L_top} is of weight 1. Given that each of the field strengths employed here are gauge covariant, this ensures gauge invariance.

Hence, the full duality invariant formulation of the theory is given by the following action
\begin{equation}
\begin{aligned}
\mc{L}=&-\fr12 \,e \, \hat{R}[g,\F] + \fr{1}{4\a_d} \,e \, g{}^{\m\n}\, \mc{D}_\m \M_{MN} \,\mc{D}_\n \M{}^{MN}-\fr{e}{2\cdot 3!}\F_{\m\n\r\, i}\M^{ij}\F^{\m\n\r}{}_j\\
&-\fr{e}{4}\F_{\m\n}{}^M\F^{\m\n N} \M_{MN}-eV+\mc{L}_{top},
\end{aligned}
\end{equation}
with the topological Lagrangian given by \eqref{L_top}. In addition one has to impose the following self-duality condition by hands
\begin{equation}
\label{self-dual}
\F_{\m\n\r\, i}=-\fr{1}{3!}e^{-1}\e_{\m\n\r\s\k\l}\,\h_{ij}\M^{jk}\F^{\s\k\l}{}_k.
\end{equation}
Note, that here we use the fully-covariant field strengths. It is important to mention, that equations of motion for the 3-form potential give this self-duality relation only under the derivative $\g^{iMN}\dt_N$. To return to the true action and the $GL(5)$ formulation one has to fix the form of the $SO(5,5)$ invariant matrix $\M^{ij}$ as in \eqref{M_fix}.

\subsection{The invariant potential}

Scalar fields of the theory are encoded in the generalised metric $\M_{MN}$, which transforms as a 
tensor of weight $\l(\M)=0$. Recall the expression for the transformation law of a tensor of weight $\l$:
\begin{equation}
\d_\L T{}^M=\L^N\dt_N T{}^M-\a_d\, \PP{}^M{}_N{}^K{}_L\, \dt_K \L^L T{}^N+\l (\dt_K \L^K)T{}^M.
\end{equation}
Although a generalised vector on extended space transforming as \eqref{Lie} has the weight $\l=\b_d$, 
in this section we will need a more general class of fields with a different weight.

Now we would like to construct a potential for the scalar fields $\M_{MN}$ that is gauge invariant and 
includes derivatives with respect to $\XX{}^M$ of the generalised metric as well as the external 
metric $g_{\m\n}$ and its determinant $g=\det g_{\m\n}$. The desired expression turns out to be:
\begin{equation}
\label{V}
\begin{aligned}
V=&\ -\fr{1}{4\a_d}\M{}^{MN}\dt_M \M{}^{KL}\dt_N \M_{KL}+\fr12 \M{}^{MN}\dt_M \M{}^{KL}\dt_L \M_{NK}\\
&-\fr12 (g{}^{-1}\dt_M g)\dt_N \M{}^{MN}-\fr14 \M{}^{MN}(g{}^{-1}\dt_M g)(g{}^{-1}\dt_N g)-\fr14 \M{}^{MN}\dt_M 
g{}^{\m\n}\dt_{N}g_{\m\n},
\end{aligned}
\end{equation}
where the terms in the first line are precisely those of \cite{Berman:2011jh}, while the rest of the terms are 
needed to ensure gauge invariance. One should note the determinant $\sqrt{-g}$ in the action~\eqref{action}.

The most convenient way to check that the above potential is invariant under the transformations induced by generalised Lie 
derivative is to introduce a non-covariant variation:
\begin{equation}
\D_\L=\d_\L-\mc{L}_\L,
\end{equation}
which measures how much the variation $\d$ of an non-covariant expression differs from its covariant 
variation. Then it is sufficient to check only the variations of non-covariant terms, e.g. for the first 
term in the potential we have:
\begin{equation}\label{1st-term}
\begin{aligned}
&\d_\L(\M{}^{MN}\dt_M \M{}^{KL}\dt_N \M_{KL})=\\
&=\d_\L \M{}^{MN}\dt_M \M{}^{KL}\dt_N \M_{KL}+\M{}^{MN}\d_\L(\dt_M \M{}^{KL})\dt_N \M_{KL}+\M{}^{MN}\dt_M 
\M{}^{KL}\d_\L(\dt_N \M_{KL})\\
&=\mc{L}_\L(\M{}^{MN}\dt_M \M{}^{KL}\dt_N \M_{KL})+\M{}^{MN}\D_\L(\dt_M \M{}^{KL})\dt_N \M_{KL}\\
&+\M{}^{MN}\dt_M \M{}^{KL}\D_\L(\dt_N \M_{KL}).
\end{aligned}
\end{equation}
The first term in the last line above automatically gives a gauge-covariant expression and we are 
left only with the last two terms. 

Let us now explicitly calculate the non-covariant variation of the term $\dt_M \M{}^{KL}$ and then list 
the corresponding variations for the other relevant expressions. Thus, we write:
\begin{equation}
\begin{aligned}
\d_\L(\dt_M \M{}^{KL})=&\ \dt_M\Big(\L^N\dt_N \M{}^{KL}-2\, \a_d\, \PP{}^P{}_Q{}^{(K}{}_N \, 
\M{}^{L)N}\dt_P\L^Q\Big),\\
\mc L_\L(\dt_M \M{}^{KL})=&\ \L^N\dt_N\dt_M\M{}^{KL}-2\, \a_d\, \PP{}^P{}_Q{}^{(K}{}_{N}\, \dt_M\M{}^{L)N}\dt_P\L^Q\\
					&+\a_d\, \PP{}^P{}_Q{}^{N}{}_{M}\, \dt_P\L^Q\dt_N \M{}^{KL}+\l(\dt 
\M)\, \dt_N\L^N \dt_M\M{}^{KL},
\end{aligned}
\end{equation}
where we added a non-zero weight for $\dt_M \M{}^{KL}$. We simplify the last line by using the section 
constraint and by setting the weight to be $\l(\dt \M)=-\b_d$, which leads to:
\begin{equation}
\D_\L (\dt_M \M{}^{KL})=-2\,\a_d\, \PP{}^R{}_P{}^{(K}{}_Q\, \M{}^{L)Q}\dt_{MR}\L^P.
\end{equation}
This choice of the weight $\l(\dt M)$ can be motivated by the fact that a geometric generalised 
vector, i.e. an object transforming as \eqref{Lie}, has a weight $\b_d$. Hence, a derivative with 
respect to the coordinate $\XX{}^M$ should add a weight $-\b_d$ to any expression. 

Following the same steps one constructs non-covariant variations for the other relevant expressions 
and obtains:
\begin{equation}
\begin{aligned}
\D_\L (\dt_N \M_{KL})=&+2\,\a_d\, \PP{}^R{}_P{}^{Q}{}_{(K}\, \M_{L)Q}\dt_{NR}\L^P,\\
\D_\L (g{}^{-1}\dt_M g)=&\ 2\, d\, \b_d\,  \dt_{MN}\L^N,\\
\D_\L (\dt_M g{}^{\m\n})=&- 2\,\b_d\, \dt_{MN}\L^N g{}^{\m\n},\\
\D_\L (\dt_M g_{\m\n})=&\ 2\, \b_d\, \dt_{MN}\L^N g_{\m\n}.
\end{aligned}
\end{equation}
Note that the weight $\l(e{}^{\ba}_\m)=\b_d$ for the vielbein derived  in the previous section implies the 
following values:
\begin{equation}
\label{weights}
\begin{aligned}
\l(g{}^{-1}\dt_M g)=- \b_d, && \l(\dt_M g{}^{\m\n})=-3\b_d, && \l(\dt_Mg_{\m\n})=\b_d.
\end{aligned}
\end{equation}
With these conventions the total weight of each term in the potential together with the prefactor of 
$e=\det e{}^{\ba}_\m$ is precisely 1.

Putting all of this together we get for the variation~\eqref{1st-term} of the first term in the potential:
\begin{equation}
\label{dV1}
\begin{aligned}
\d_\L\Big(-\fr{e}{4\a_d}\,\M{}^{MN}\,\dt_M \M{}^{KL}\,\dt_N \M_{KL}\Big)\to &\ 
e\,\M{}^{MN}\,\PP{}^P{}_Q{}^{(K}{}_R\,\M{}^{L)R}\,\dt_N \M_{KL}\,\dt_{MP}\L^Q \\
=& \ e\, \M{}^{MN}\M{}^{KL}\,\dt_M \M_{KP}\,\dt_{LN}\L^P.
\end{aligned}
\end{equation}
In the second line we used the fact that the matrix $\M{}^{MN}$ parameterises the coset $G/K$ with $G$ being the U-duality group. Then one is able to construct a current
\begin{equation}
\label{J}
(J_M){}^P{}_Q:= \M{}^{PR}\dt_M \M_{RQ},
\end{equation}
that belongs to the algebra $\mathfrak{g}$ of the group $G$ and is invariant under the action of the 
projector on the adjoint. Hence, we write
\begin{equation}
\label{PJ}
\PP{}^P{}_Q{}^{K}{}_L (J_N){}^L{}_K=(J_N){}^P{}_Q.
\end{equation}

For the non-covariant part of the variation of the second term in the potential we obtain
\begin{equation}
\begin{aligned}
&\d_\L\Big(\fr e2\, \M{}^{MN}\dt_M \M{}^{KL}\dt_L \M_{NK}\Big)\to \\
& \to -\fr{e}{2} \, \a_d \Big( \PP{}^R{}_P{}^{(K}{}_Q\, \M{}^{L)Q}\M{}^{MN}\dt_L \M_{NK}\dt_{MR}\L^P-
\PP{}^R{}_P{}^Q{}_{(N}\, \M_{K)Q}\M{}^{MN}\dt_M \M{}^{KL}\dt_{LR}\L^P\Big)\\
&=-e\, \a_d\, \M{}^{MN}\PP{}^R{}_P{}^Q{}_{N}\,(J_M){}^L{}_Q\, \dt_{LR}\L^P + e\, \b_d\, \dt_K \M{}^{KL}\dt_{LP}\L^P + e \, 
\dt_P \M{}^{KL}\,\dt_{KL}L{}^{P},
\end{aligned}
\end{equation}
where the section condition was used in the third line. To cancel the variation of the first term in 
the potential \eqref{dV1} one has to modify the first term in the last line above. Using the 
property \eqref{PJ} of the current me rewrite this term as
\begin{equation}
\begin{aligned}
e\, \a_d\, \M{}^{MN}\,\PP{}^R{}_P{}^Q{}_{N}\,(J_M){}^L{}_Q\, \dt_{LR}\L^P=e\, \a_d\, 
\M{}^{MN}\,\PP{}^R{}_P{}^Q{}_{N}\,\PP{}^Q{}_L{}^U{}_{V}\,(J_M){}^V{}_U\,\dt_{LR}\L^P.
\end{aligned}
\end{equation}
Next, expressing the projectors back in terms of the tensor $Y{}^{MN}_{KL}$ and using the invariance 
condition in the first line of \eqref{rel} we obtain for this term:
\begin{equation}
\begin{aligned}
& e\, \b_d\, \M{}^{MN}\,\PP{}^L{}_N{}^U{}_V\, (J_M){}^V{}_U\, \dt_{LP}\L^P + e\,\M{}^{MN}\,\PP{}^L{}_P{}^U{}_V\, 
(J_M){}^V{}_U\,\dt_{LN}\L^P\\
&=-e\,\b_d\, \dt_M M{}^{LM}\,\dt_{LP}\L^P + e\, \M{}^{MN} \M{}^{LK}\,\dt_M \M_{KP}\,\dt_{LN}\L^P.
\end{aligned}
\end{equation}
Hence, in total for the non-covariant part of the variation of the second term in the potential we 
have:
\begin{equation}
\label{dV2}
\begin{aligned}
& \d_\L\Big(\fr e2\, \M{}^{MN}\,\dt_M \M{}^{KL}\dt_L \M_{NK}\Big)\to \\
&\to -e\, \M{}^{MN}\M{}^{LK}\,\dt_M M_{KP}\,\dt_{LN}\L^P + 2\,\b_d \,\dt_K \M{}^{KL}\,\dt_{LP}\L^P + e \, \dt_P 
\M{}^{KL}\,\dt_{KL}\L^{P},
\end{aligned}
\end{equation}
and the variation \eqref{dV1} is successfully cancelled. The remaining terms linear in $\dt \M$ are 
cancelled with the terms coming from the second line in the potential~\eqref{V}. 

Indeed, consider the contraction
\begin{equation}
\D_\L (\dt_N \M{}^{MN})=-(2\b_d+1)\, \dt_{NP}\L^P \M{}^{MN} - \M{}^{NK}\,\dt_{NK}\L^M,
\end{equation}
where the section constraint was taken into account. Then, the non-covariant variations of the terms 
3, 4 and 5 in the potential can be written as
\begin{equation}
\begin{aligned}
\D_\L(3)=&- d\,\b_d\, e\, \dt_{MP}\L^P\, \dt_N \M{}^{MN} + 2\,\b_d\, \dt_M\, e\, \dt_{NP}\L^P\, \M{}^{MN} + \M{}^{NK}\,\dt_M 
e\, \dt_{NK}\L^M,\\
\D_\L(4)=&- 2\,d\, \b_d\, \M{}^{MN}\,\dt_M e\, \dt_{NP}\L^P,\\
\D_\L(5)=& \ 2\, \b_d\, \M{}^{MN}\, \dt_M e\, \dt_{NP}\L^P.
\end{aligned}
\end{equation}
Altogether, combining these with the remaining pieces from \eqref{dV2} we obtain for the total 
variation 
\begin{equation}
\begin{aligned}
\d_\L(e\, V ) = &\ \dt_N\,(e\,\L^N V)+e\, \D_\L V\\
=&\ \dt_N\,(e\,\L^N V) - e\, \dt_{MP}\L^P\, \dt_N \M{}^{MN} + e\, \dt_P\M{}^{KL}\,\dt_{KL}\L^P\\
&-\dt_M e\, \M{}^{MN}\,\dt_{NP}\L^P  + \M{}^{KL}\,\dt_P e\, \dt_{KL}\L^P\\
=& \ \dt_N\big(e\,\L^N V - e\,\dt_{PQ}\L^P\, M{}^{QN} + e\, \M{}^{KL}\,\dt_{KL}\L^N\big)\to 0,
\end{aligned}
\end{equation}
where we used the identity $g{}^{-1}\dt_M g = 2\,e{}^{-1}\,\dt_M e$. 

Hence, it has been explicitly shown that the potential for the scalar fields \eqref{V} is invariant 
under the transformations induced by the generalised Lie derivative up to boundary terms, which drop from 
the corresponding action. Remarkably, all the coefficients are fixed by the gauge invariance, up to 
an overall prefactor. Moreover, weight counting for the terms in the potential together with the 
invariance of the Einstein-Hilbert term give the correct pattern for $\b_d$. It is interesting to note, that 
although the invariance condition for the $Y$-tensor looks differently for the $E_{6(6)}$, the 
scalar potential is given by the same expression \eqref{V}. 

Finally, in precise analogy with the $E_{7(7)}$ case~\cite{Hohm:2013uia}, to compare with the 
previous results~\cite{Berman:2011jh} for the potential for $\M_{MN}$ one uses the truncation 
$g_{\m\n}=e{}^{2\D}\h_{\m\n}$ and rescales the generalised metric as $\M_{MN}\to e{}^{\g \D} \M_{MN}$. To 
ensure the U-duality invariance the field $\D=\D(\XX)$ must be an independent degree of freedom.

\subsection{External D=5+1 diffeomorphisms}

We have seen that invariance of the Lagrangian with respect to gauge 
transformations generated by the generalised Lie derivative fixes the relative coefficients of different terms inside the potential. Same as in the EFT's for the other duality groups, the 
relative coefficients of different terms within~\eqref{action} are fixed by imposing invariance with respect to the external 
diffeomorphisms. 
For a diffeomorphism generated by a parameter $\x{}^\m$ 
that does not depend on the extended coordinates $\XX{}^M$, each term in the Lagrangian is manifestly 
invariant. However, the situation becomes more subtle if one considers a general dependence of the 
parameter on extended coordinates. In close analogy with the other EFT's we consider the following 
transformations:
\begin{equation}
\label{diffeo}
 \begin{aligned}
  \d e{}^{\ba}_\m &=\x{}^\m \mc{D}_\n e{}^{\ba}_\m+\mc{D}_\m \x{}^\n e{}^{\ba}_\n=L_\x{}^\DD e{}^{\ba}_\m,\\
  \d \M_{MN} &= \x{}^\m \mc{D}_\m \M_{MN}=L_\x{}^\DD \M_{MN},\\
   \d A{}^M_\m &= \x{}^\n \F_{\n\m}+ \M{}^{MN}g_{\m\n}\dt_N \x{}^\n=L_\x{}^\DD A_\m^M+\ldots,\\
   \D B_{\m\n\, i}&=\x{}^\r \mc{G}_{\r\m\n\, i}=L_\x{}^\DD B_{\m\n\, i}+\ldots,\\
   \D C_{\m\n\r\, N}&= \fr{e}{4\k} \e_{\m\n\r\s\k\l}\x{}^\s \F{}^{\k\l\, M}\M_{MN}.
 \end{aligned}
\end{equation}
Here $L_\x{}^\DD$ denotes the conventional Lie derivative along $\x{}^\m$ built from the 
covariantised derivatives $\DD_\m$. Transformation of the 3-form potential is required to be of 
this particular form by invariance of 
the Lagrangian. Note however, that this is equal to the conventional form $\D C_{\m\n\r\, N}=\x{}^\s 
\F_{\s\m\n\r\, M}$ on the equations of motion of the ``magnetic'' 2-form potential \eqref{dual_C} for 
any $\k$.

In what follows we will focus mainly on the terms that contain the derivative $\dt_M \x{}^\m$, 
referring to them as new terms. By contrast, cancellation of the 
other contributions works in a way similar to the maximal gauged supergravity and hence does not require a 
detailed analysis.

Let us start first with transformation of the kinetic term for the scalar fields $\mc{M}_{MN}$, 
whose cancellation with the kinetic term for vector fields is universal. Hence, we write
\begin{equation}
 \begin{aligned}
   \d_\x(\DD_\m \M_{MN})&= 
L_\x{}^{\DD}(\DD_\m\M_{MN})+2\,\a_d\,\M_{P(M}\PP_{N)}{}^P{}_R{}^S\,\mc{F}_{\m\n}{}^R\,\dt_S\x{}^\n\\
&-\M{}^{KP}\dt_K \M_{MN}\,\dt_P \x{}^\n g_{\m\n}-2\,\a_d\, \M_{P(M}\PP_{N)}{}^P{}_R{}^S\,
\dt_S(\M{}^{RQ}g_{\m\n}\,\dt_Q \x{}^\n). 
\end{aligned}
\end{equation}
Substituting this into the variation of the kinetic term for scalars and keeping only the 
relevant terms we obtain:
\begin{equation}
\label{kin_scal}
 \begin{aligned}
   \fr{1}{4\a_d}\,\d_\x(e\,g{}^{\m\n}\DD_\m \M_{MN}\,\DD_\n \M{}^{MN})& = e\,g{}^{\m\r}\DD_\r \M{}^{MN} 
\M_{NK}\,\mc{F}_{\m\n}{}^K\dt_M\x{}^\n\\
& + e\,(\M_{NL}\dt_M\M{}^{LK}-\fr{1}{2\a_d}\M{}^{KL}\dt_L \M_{MN} )\DD_\m \M{}^{MN}\dt_K \x{}^\m\\
&+\ldots.
 \end{aligned}
\end{equation}
Here the dots denote the omitted part of the variation which is not relevant for setting up the relative coefficients between 
the terms of~\eqref{action}. The first term above will be cancelled by a corresponding contribution from the variation of 
the kinetic term of the 1-form potential. 

In order to cancel the second term in the expression above we consider the variation of the scalar potential 
$V$, which enters the Lagrangian with negative sign. Again, following only the most indicative terms 
we write (cf. \cite{Hohm:2013vpa}):
\begin{equation}
 \begin{aligned}
   \d_\x V&=\d_\x\left(\fr12\, \M_{NL}\,\dt_M \M{}^{LK}-\fr{1}{4\a_d}\,\M{}^{KL}\,\dt_L\M_{MN}\right)\dt_K \M{}^{MN}+\ldots\\
   &=\left(\M_{NL}\,\dt_M \M{}^{LK}-\fr{1}{2\a_d}\,\M{}^{KL}\,\dt_L\M_{MN}\right)\DD_\m  \M{}^{MN} \dt_K \x{}^\m + \ldots.
 \end{aligned}
\end{equation}
We observe that this variation successfully cancels the variation~\eqref{kin_scal}, in line with what appears to be a 
common behaviour of every EFT.

To see the other cancellations, let us turn to the vector-tensor sector of the model. The 
corresponding variation is given in the duality covariant form~\eqref{delta_L1}. In what follows 
we will drop variations of the density $e$ and the external metric $g{}^{\m\n}$, which as usual 
complete the variations of the other terms to full derivatives. Hence, for the terms in~\eqref{delta_L1} we have:
\begin{equation}
\label{all_var1}
 \begin{aligned}
   (1)&=-\fr12\,e\,\M_{MN}\,\F{}^{\m\n\, M}\delta \F_{\m\n}{}^N,\\
   (2)&=-\fr{\k}{3!}\,\e{}^{\m\n\r\s\k\l}\,\GG_{\m\n\r\, i}\DD_{\s}\D B_{\k\l\, j}\,\h{}^{ij},\\
   (3)&=\fr{\sqrt{2}\,\k}{8}\,\e{}^{\m\n\r\s\k\l}\,\F_{\m\n}{}^M \g{}^{i}{}_{MN}\F_{\r\s}{}^N\,\D B_{\k\l\, i},\\
   (4)&=-\fr{\sqrt{2}\,\k}{3!}\,\e{}^{\m\n\r\s\k\l}\,\GG_{\m\n\r\, i}\,\g{}^{i}{}_{MN}\F_{\s\k}{}^M\,\D A_\l{}^N,\\
   (5)&=-\fr{\sqrt{2}\,\k}{3\cdot 4!}\,\e{}^{\m\n\r\s\k\l}\,(\GG_{\m\n\r\, i}-\F_{\m\n\r\, i})\,\g{}^{i\,MN}\,\dt_M \D C_{\s\k\l\, N},
 \end{aligned}
\end{equation}
where in the last term we have traded the field $C_{\m\n\r}{}^{M,KL}$ for $C_{\m\n\r\, M}$ 
for convenience. 

Let us start with the terms (3) and (4), which upon substitution of the explicit expressions for the 
variation~\eqref{diffeo} give:
\begin{equation}
\label{3+4}
 \begin{aligned}
&(3)+(4)=\\
=&-\fr{\sqrt{2}\,\k}{3!}\,\e{}^{\m\n\r\s\k\l}\,\GG_{\m\n\r\, i}\g{}^{i}{}_{MN}\,\F_{\s\k}{}^M\F_{\f\l}{}^N\x{}^\f- 
\fr {\sqrt{2}\,\k}{3!}\,\e{}^{\m\n\r\s\k\l}\,\GG_{\m\n\r\, i}\g{}^{i}{}_{MN}\,\F_{\s\k}{}^M \M{}^{NK}g_{\l\f}\dt_K \x{}^\f \\
&+\fr{\sqrt{2}\,\k}{8}\,\e{}^{\m\n\r\s\k\l}\,\x{}^\f\GG_{\f\k\l\, i}\g{}^{i}{}_{MN}\,\F_{\m\n}{}^M \F_{\r\s}{}^N.
 \end{aligned}
\end{equation}
The first and the last terms together can be organised into an expression with seven indices 
$\{\m\n\r\s\k\l\f\}$ antisymmetrised and hence vanish.

To see further cancellations  consider the variation of the 2-form field strength:
\begin{equation}
 \begin{aligned}
  \d_\x \F_{\m\n}{}^M = &\ 2\,\DD_{[\m}\D A_{\n]}{}^M-\fr{1}{2\sqrt{2}}\,\g{}^{i\,MN}\,\dt_N B_{\m\n\, i}\\
  =&\ 2\, \DD_{[\m} (\x{}^\r \F_{|\r|\n]}{}^M) + 2 \,\DD_{[\m}(\M{}^{MN}\,g_{\n]\r}\,\dt_N 
\x{}^\r) -\fr{1}{2\sqrt{2}}\,\dt_N(\x{}^\r\, \GG_{\m\n\r\, i})\g{}^{i\,MN}\\
=& \ L_\x{}^\DD \F_{\m\n}{}^M-\fr{1}{2\sqrt{2}}\,\dt_N\Big(\GG_{\m\n\r\, i}-\F_{\m\n\r\, i}\Big)\x{}^\r 
\g{}^{i\,MN}-\fr{1}{2\sqrt{2}}\,\dt_N \x{}^\r\, \GG_{\m\n\r\, i}\g{}^{i\,MN}\\
& + 2\,\DD_{[\m}(\M{}^{MN}\,g_{\n]\r}\,\dt_N \x{}^\r),
 \end{aligned}
\end{equation}
where we have used the Bianchi identity for the field $\F_{\m\n}{}^M$ to organise the conventional Lie 
derivatives $L_\x{}^\DD$ everywhere. One should note the remark at 
the end of the section~\ref{Der}. The last term in the variation above being substituted 
into (1) cancels the corresponding term coming from variation of the modified Einstein-Hilbert term 
precisely in the same way as it takes place in the other EFT's. The term $L_\x{}^\DD \F_{\m\n}{}^M$ above forms a full derivative together with the variation of the determinant $e$ 
and the generalised metric $\M_{MN}$. The remaining piece in the variation (1) together with (5) 
gives:
\begin{equation}
\label{1+5}
 \begin{aligned}
  &(1)+(5)=\\
  =& \fr{1}{4\sqrt{2}}\,e\,\M_{MN}\F{}^{\m\n\, M}\,\dt_K\big(\GG_{\m\n\r\, i}-\F_{\m\n\r\, i}\big)\,\x{}^\r 
\g{}^{i\,NK}+\fr{e}{4\sqrt{2}}\,\M_{MN}\F{}^{\m\n\, M} \GG_{\m\n\r\, i}\g{}^{i\,NK}\,\dt_K \x{}^\r \\
&-\fr{\sqrt{2}\,\k}{16}\,\big(\GG_{\m\n\r\, i}-\F_{\m\n\r\, i}\big)\g{}^{i\,MN}\,\dt_M  \big(e\, \M_{NK}\,\x{}^\r\,\F{}^{\m\n\, K}\big),
 \end{aligned}
\end{equation}
where we have used the explicit from of the variation $\D_\x C_{\m\n\r\, M}$ and contracted two 
epsilon tensors. Observe that the first and the last terms above cancel each other 
off-shell.

The remaining term above cancels with the corresponding piece in \eqref{3+4} if one chooses $\k=1/2$ and takes into account the  self-duality condition for the field strengths $\GG_{\m\n\r\, i}$ dressed up with the scalar matrix \cite{Cremmer:1997ct}:
\begin{equation}
\fr{1}{3!}\,\e{}^{\m\n\r\s\k\l}\,\GG_{\s\k\l\, i}\,\g{}^i{}_{MN}\M{}^{NK}=e\, \M_{MN}\GG{}^{\m\n\r}{}_i\,\g{}^{i\,NK}.
\end{equation}

It is important to mention here that in the case of $D=6$ maximal gauged supergravity the factor $\k$ remains undetermined unless one considers supersymmetry invariance. The novel feature of the EFT approach is that it is fixed at the level of bosonic equations of motion.

Finally, the term (2) works in the same way as for the $D=6$ maximal gauged supergravity,
forming a full derivative together with the variation of the determinant $e$ and the scalar matrix 
$K_{mn}$.

\section{Embeddings of  D=11 and Type IIB supergavity}
\label{Embed}

The coordinate space of the $SO(5,5)$ Exceptional Field Theory is parameterised by six external 
coordinates $x{}^\m$ and 16 extended coordinates $\XX{}^M$. Dynamics along the latter is restricted by 
the section condition
\begin{equation}
 \g{}^{i\,MN}\dt_M\bullet \dt_N \bullet =0.
\end{equation}
In this section we consider two solutions of this equation that break the $SO(5,5)$ duality group 
to $GL(5) \simeq SL(5) \times GL(1)$ and $GL(4)\times SL(2)$. The corresponding split of the field content of EFT 
gives the field content of $D=11$ and Type IIB supergravities respectively. In the latter case one 
finds a manifest $SL(2)$ covariant formulation. 

Let us start with the decomposition with respect to the $SL(5)\times GL(1)$ subalgebra. Since this contains a $GL(1)$ subgroup this decomposition is performed by removing a node from the Dynkin diagram for $SO(5,5)$:\footnote{All branching rules provided in this section were obtained by using the Mathematica package LieART \cite{Feger:2012la}. This reference is also recommended for theoretical background on subalgebra decomposition and branching rules, and for further references.}
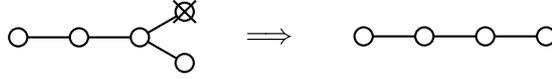
\begin{center}
\parbox{3cm}{
  \begin{tikzpicture}[scale=.4]
    \foreach \x in {2,...,4}
    \draw[xshift=\x cm,thick] (\x cm,0) circle (.3cm);
    \draw[xshift=8 cm,thick] (30: 17 mm) circle (.3cm);
    \draw[xshift=8 cm,thick] (-30: 17 mm) circle (.3cm);
    \draw[xshift=8 cm,thick] (30: 17 mm) node[cross,rotate=0] {};
    \foreach \y in {2.15,...,3.15}
    \draw[xshift=\y cm,thick] (\y cm,0) -- +(1.4 cm,0);
    \draw[xshift=8 cm,thick] (30: 3 mm) -- (30: 14 mm);
    \draw[xshift=8 cm,thick] (-30: 3 mm) -- (-30: 14 mm);
  \end{tikzpicture}
  }
  $\Longrightarrow$
  \parbox{4cm}{
  \begin{center}
    \begin{tikzpicture}[scale=.4]
      \foreach \x in {1,...,4}
      \draw[xshift=\x cm,thick] (\x cm,0) circle (.3cm);
      \foreach \y in {1.15,...,3.15}
      \draw[xshift=\y cm,thick] (\y cm,0) -- +(1.4 cm,0);
    \end{tikzpicture}
  \end{center}
  }
\end{center}
The corresponding branching rules for the relevant representations take the following form
\begin{equation}
 \begin{aligned}
   &\bf 16 && \longrightarrow &&  \bar{5}_{+3}  \oplus 10_{-1} \oplus 1_{-5},\\
   &\bf 10 && \longrightarrow &&  5_{+2} \oplus \bar{5}_{-2},
 \end{aligned}
\end{equation}
where the subscript denotes weight with respect to $GL(1)$ rescalings. Using the decomposition of $\bf 16$ 
we have for the coordinate $\XX{}^M$:
\begin{equation}
 \begin{aligned}
  & \{\XX{}^M\} && \longrightarrow && \{ x{}^m,\, y_{mn}, \, z_{mnpqr} \},
 \end{aligned}
\end{equation}
where $x{}^m$ is the conventional geometric coordinate, while $y_{mn}$ and $z_{mnpqr}$ correspond to the
winding modes of the M2- and M5-branes. To solve the section condition one leaves only the dependence 
of the five coordinates $x{}^m$ that restores the eleven-dimensional spacetime of the $D=11$ 
supergravity.

In the on-shell formulation equations of motion for the 3-form fields $C_{\m\n\r\, M}$ give the self-duality relation for the 2-form potentials leaving only five of ten. Hence, for the $p$-forms we have the following:
\begin{equation}
 \begin{aligned}
   & A_\m^M && \longrightarrow && A_\m^m, \, A_{\m\, mn}, \, A_\m;\\
   & B_{\m\n\, i} && \longrightarrow && B_{\m\n\, m}.
 \end{aligned}
\end{equation}
This nicely fits into the decomposition of eleven-dimensional fields under the split $11=6+5$, 
that is (see \eqref{split}):
\begin{equation}
 \begin{aligned}
   & G_{\hat{\tt{M}}\hat{\tt{N}}} && \longrightarrow && g_{\m\n}, \, A_{\m}^m, \, \f_{mn};\\
   & C_{\hat{\tt{M}}\hat{\tt{N}}\hat{\tt{K}}} && \longrightarrow && C_{\m\n\r}, \, B_{\m\n\, m}, \, A_{\m\, mn}, \, 
\f_{mnp}.
 \end{aligned}
\end{equation}
Upon dualizing the 3-form field $C_{\m\n\r}$ one identifies all the 
1-forms here. The five 2-forms are identified with five electric 2-form potentials $B_{\m\n\, m}$ of 
the EFT. Note, that one is free to choose the five electric forms among ten $B_{\m\n\, i}$ by 
choosing an appropriate U-duality frame. If one works off-shell and keeps the 3-forms one has to keep the magnetic 2-forms as well and identify these to the fields coming from the magnetic 6-form potential of 11-dimensional supergravity. However, since we are working in the true action formalism it is more consistent to keep the discussion essentially on-shell. 

The scalar matrix $\M_{MN}$ is built from the coset representative $\mc{V}_M{}^{\a\dot{\a}}$, which is an element of $SO(5,5)$. The adjoint representation $\bf{45}$ of $SO(5,5)$ is decomposed under $SL(5)$ as follows
\begin{equation}
\begin{aligned}
 \bf{45} && \longrightarrow && 1_0+24_0+10_{+4}+\overline{10}_{-4}.
\end{aligned}
\end{equation}
The compact subgroup of the last two terms correspond to generators of one of the $SO(5)$ in the local subgroup $SO(5)\times SO(5)$ and hence drop. The other $SO(5)$ appears as the compact subgroup of the $SL(5)$ generators given by $24_0$ and should be dropped as well. The remaining 25 degrees of freedom correspond to the symmetric matrix $\f_{mn}$ and the 3-form $\f_{mnk}$.

Decomposition of $SO(5,5)$ with respect to a $GL(4)$ that is not a subgroup of the $GL(5)$ above is performed by adding the most negative root to the Dynkin diagram. The resulting diagram becomes linearly dependent and decomposes into a sum. Hence, for algebras in the $D_n$  class we have 
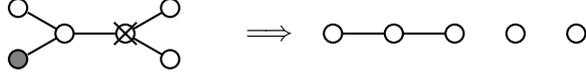
\begin{center}
\parbox{3cm}{
  \begin{tikzpicture}[scale=.4]
    \foreach \x in {3,...,4}
    \draw[xshift=\x cm,thick] (\x cm,0) circle (.3cm);
    \draw[xshift=8 cm,thick] (30: 17 mm) circle (.3cm);
    \draw[xshift=8 cm,thick] (-30: 17 mm) circle (.3cm);
    \draw[xshift=3 cm,thick] (30: 17 mm) circle (.3cm);
    \draw[xshift=3 cm,thick,fill=gray] (-30: 17 mm) circle (.3cm);
    \draw[xshift=8 cm,thick] (0,0) node[cross,rotate=0] {};
    \foreach \y in {3.15,...,3.15}
    \draw[xshift=\y cm,thick] (\y cm,0) -- +(1.4 cm,0);
    \draw[xshift=8 cm,thick] (30: 3 mm) -- (30: 14 mm);
    \draw[xshift=8 cm,thick] (-30: 3 mm) -- (-30: 14 mm);
    \draw[xshift=6 cm,thick] (150: 3 mm) -- (150: 14 mm);
    \draw[xshift=6 cm,thick] (-150: 3 mm) -- (-150: 14 mm);
  \end{tikzpicture}
  }
  $\Longrightarrow$
  \parbox{4cm}{
  \begin{center}
    \begin{tikzpicture}[scale=.4]
      \foreach \x in {2,...,6}
      \draw[xshift=\x cm,thick] (\x cm,0) circle (.3cm);
      \foreach \y in {2.15,...,3.15}
      \draw[xshift=\y cm,thick] (\y cm,0) -- +(1.4 cm,0);
    \end{tikzpicture}
  \end{center}
  }
\end{center}
where the added root is denoted by the grey circle. Under this procedure the algebra $SO(5,5)$ is decomposed as
\begin{equation}
SO(5,5) \hookleftarrow SL(4)\oplus SL(2) \oplus SL(2).
\end{equation}
To identify geometric and winding coordinates among $\XX{}^M$ one writes the corresponding branching rule for the $\bf 16 $ representation
\begin{equation}
 \begin{aligned}
  & \bf 16 && \longrightarrow && (4,1,2) \oplus (\bar{4},2,1).
 \end{aligned}
\end{equation}
One has here two pairs of four coordinates each pair transforming under one of the $SL(2)$ algebras in the decomposition. We identify the representation $(\bar{4},2,1)$ with the doublet of winding coordinates $ y_{\un{m}\, \hat{\a}}$ corresponding to the fundamental F1-string and the D1-brane. The corresponding $SL(2)$ is then identified with S-duality group of Type IIB theory. 

The remaining $SL(2)$-doublet $(4,1,2)$ is composed of the geometric coordinates $x{}^{\un{m}}$ corresponding to translational modes and the coordinates $z_{\un{mnr}}$ corresponding to windings of the D3-brane. This explicit choice breaks the $SL(2)$ symmetry leaving only its $GL(1)$ subgroup. Hence, we have the following decomposition for extended coordinates
\begin{equation}
 \begin{aligned}
  & \{\XX{}^M \}&& \longrightarrow && \{x{}^{\un{m}}, y_{\un{m}\, \hat{\a}}, z_{\un{mnr}}\}.
 \end{aligned}
\end{equation}

Upon this choice of the solution of section condition one considers the embedding $GL(4)\times SL(2) \hookrightarrow SO(5,5)$ and the corresponding branching rules read
\begin{equation}
 \begin{aligned}
  & \bf 16 && \longrightarrow && (4,1)_{+1} \oplus (\bar{4},2)_0 \oplus (4,1)_{-1};\\
  & \bf 10 && \longrightarrow && (1,2)_{+1} \oplus (6,1)_0 \oplus (1,2)_{-1},
 \end{aligned}
\end{equation}
where the subscript denotes weight with respect to the $GL(1)$.

By construction it is manifest that the $GL(4)$ group here is not a subgroup of the $GL(5)$ group above. Such a case would correspond to Type IIA supergravity that is a reduction of $D=11$ supergravity on a circle. A nice explicit example of the relation between Type IIA and Type IIB supergravities in the $O(3,3)$ formulation coming from reduction of the $SL(5)$ covariant 
field theory is provided in~\cite{Thompson:2011uw}.

Field content of Type IIB supergravity is decomposed as follows:
\begin{equation}
 \begin{aligned}
   & G_{\tt{M}\tt{N}} && \longrightarrow && g_{\m\n}, \, A_{\m}^{\un{m}}, \, \f_{\un{mn}};\\
   & C_{\hat{\a}} && \longrightarrow && \f_{\hat{\a}};\\
   & B_{\tt{M}\tt{N} \hat{\a}} && \longrightarrow && B_{\m\n\,\hat{\a}}, \, A_{\m\, \un{m}\, \hat{\a}}, \, \f_{\un{mn}\, \hat{\a}};\\
   & C_{\tt{M}\tt{N}\tt{K}\tt{L}} && \longrightarrow &&  B_{\m\n\, \un{mn}}, \, A_{\m\, \un{mnr} }, \, \f_{\un{mnrs}}, \, C_{\m\n\r\s}, \, C_{\m\n\r\, \un{m}}. 
 \end{aligned}
\end{equation}
The last two fields and a half of d.o.f's of the 2-form field in the last line should be dropped due to the self-duality condition in 10 dimensions.

The representation $\bf 45$ parameterised by the generalised vielbein $\mc{V}$ under the algebra decomposition goes according to the following rule
\begin{equation}
\begin{aligned}
 \bf{45} && \longrightarrow && (1,1)_{+2}+(1,1)_0+(1,1)_{-2}+(2,6)_{+1}+(2,6)_{-1}+(3,1)_0+(1,15)_0.
\end{aligned}
\end{equation}
We see, that the $SO(5)\times SO(5)$ subalgebra is broken and one can see here only the $O(4)\times O(4)$ generators corresponding to the T-duality coset $O(n,n)/O(n)\times O(n)$. As in the previous case, one of these $O(4)$ appears as a compact part of  $(2,6)_1\oplus (2,6)_{-1}$ of $SL(4)$ and the other comes from $(1,15)_0$.

On the level of fields, the scalar matrix $\M_{MN}$ is composed of the 25 scalars in the usual way~\cite{Cremmer:1998em}:
\begin{equation}
\{\f_{\hat{\a}},\f_{\un{mn}},\f_{\un{mn} \hat{\a}},\f_{\un{mnrs}}\} \longrightarrow \M_{MN}.
\end{equation}
The vector fields are collected according to the decomposition of the $\bf 16$:
\begin{equation}
\{A_{\m}^{\un{m}},A_{\m\, \un{m}\, \hat{\a}}, A_{\m\, \un{mnr} }\} \longrightarrow  A_\m^M.
\end{equation}
There are only five 2-form fields in the field content that correspond to the five electric 2-forms:
\begin{equation}
\{ B_{\m\n\,\hat{\a}}, B_{\m\n\, \un{mn}}\} \longrightarrow  B_{\m\n\, m} .
\end{equation}
Note that there remain only three of six 2-forms $B_{\m\n \un{mn}}$ due to the self-duality condition. Alternatively, one may switch to the so called democratic formulation of Type IIB supergravity~\cite{Bergshoeff:2001pv}, where all $p$-forms including their duals are present. In this case one has to keep the 3-form field $C_{\m\n\r\, M}$ and all the ten 2-forms.

\section{Outlook and conclusion}

The bosonic $SO(5,5)$ covariant field theory constructed here forms a link in the chain of Exceptional Field Theories with their gauge groups being the exceptional groups $E_{d(d)}$ \cite{Hohm:2013vpa,Hohm:2013uia,Hohm:2014fxa,Hohm:2015xna}. The key feature of EFT is the notion of generalised Lie derivative, which is an analogue of the conventional Lie derivative with an appropriate exceptional group instead of $GL(D)$. This transformation acts as a gauge symmetry of the theory, which is constructed in the spirit of Yang-Mills model.

We have shown how the unusual properties of the new gauge transformation such as the necessity of section condition and failure of the Jacobi identity naturally lead to tensor hierarchy. The story is kept as general as possible and can be carried over to the $SL(5)$ and $SL(2)\times SL(3)$ groups as well. One needs to do small modifications in the identities \eqref{rel} and \eqref{rel1} in order to go to the $E_6$ case (see \cite{Musaev:2013rq} for more detailed discussion of this issue).

We construct both the true action, which gives covariant equations of motion as well as all duality relation, and the pseudo-action, which is manifestly duality invariant. The true action is not invariant under the gauge transformations induced by local coordinate transformations of the extended space. The invariant pseudo-action takes the following  simple form:
\begin{equation}
\begin{aligned}
\mc{L}=&-\fr12 \,e \, \hat{R}[g,\F] + \fr{1}{4\a_d} \,e \, g{}^{\m\n}\, \mc{D}_\m \M_{MN} \,\mc{D}_\n \M{}^{MN}-\fr{e}{2\cdot 3!}\F_{\m\n\r\, i}\M^{ij}\F^{\m\n\r}{}_j\\
&-\fr{e}{4}\F_{\m\n}{}^M\F^{\m\n N}M_{MN}-eV+\mc{L}_{top}.
\end{aligned}
\end{equation}
Here, the topological Lagrangian is defined by an integral of an exact form over a non-physical seven-dimensional spacetime, whose boundary is the six-dimensional physical spacetime
\begin{equation}
\begin{aligned}
S_{top}&=\int d^6x\, d^{16}\XX\, \mc{L}_{top}\\
&=\int d^7X\, d^{16}\XX\, \left(2\,\h^{ij}\F_i\wedge \DD\F_j-\fr{1}{\sqrt{2}}\F\wedge\g^i\F\wedge \F_i\right).
\end{aligned}
\end{equation}
The pseudo-action is supplemented with the modified duality covariant Einstein-Hilbert term $\hat{R}[g,\F]$, that has the same form as in the other EFT's, and the scalar potential $V$ that governs the dynamics of the generalised metric $\M_{MN}$ in the extended space. The latter is written in the most general form as well. In addition one imposes the following self-duality condition by hands
\begin{equation}
*\!\F_i =-\h_{ij} \,\M{}^{jk}\, \F_{k}\, .
\end{equation}

We have shown that in order to have the potential invariant under duality transformations generated by $\L^M$ one has to fix the weights of the vielbein and generalised metric to be $\b_d$ and 0 respectively. This in turn fixes the value of $\b_d$ that perfectly reproduces the value needed for consistency of the algebra \cite{Berman:2012vc}. One concludes that the construction of EFT is very rigid and natural.

Gauge invariance constrains the action but leaves undetermined the relative coefficients between the Einstein-Hilbert term, the scalar potential, the kinetic term for vector fields and the action for 2-forms. We have demonstrated that all these are fixed by requiring the invariance with respect to external diffeomorphisms along $\x{}^\m=\x{}^\m(x,\XX)$. The action of external diffeomorphisms on the elementary fields of the theory is provided in~\eqref{diffeo}.

Hence, the action becomes completely fixed. Note, that this is the novel feature of EFT: normally the actions of maximal gauged supergravities become fixed only after imposing supersymmetry. The construction presented here considers only the bosonic sector of maximal supergravity in 6 dimensions. Fermions and supersymmetry can be added following the similar approach as in  \cite{Godazgar:2014nqa,Musaev:2014lna}.

The section constraint, which one has always to keep in mind, effectively restricts the dynamics in the extended space. There are two solutions of the condition that lead to theories in 11 and in 10 dimensions. These are given by embeddings of $GL(5)$ and $GL(4)\times SL(2)$ in $SO(5,5)$. We show that under the first embedding the field content of the constructed EFT perfectly fits the field content of $D=11$ supergravity, while the second embedding gives $D=10$ Type IIB supergravity with manifest $SL(2)$ symmetry. Note, that the $GL(4)$ is not a subgroup of the $GL(5)$. However, one is always allowed to do further branching with respect to the embedding $GL(4)\subset GL(5)$, which gives Type IIA supergravity. Hence, the Exceptional Field Theory construction considers $D=11$ supergravity and Type  IIB theory on the same footing, which is possible due to lack of 10-dimensional Lorentz symmetry.

Of special interest is the additional $SL(2)$ symmetry of Type IIB supergravity recovered in the EFT construction. Upon decomposition of the extended coordinates $\X{}^M$ this corresponds to rotations of the translational modes and the winding modes of the D3-branes. The authors are not familiar with literature that mentions this kind of hidden symmetry and avoid any interpretation based on such schematic derivation. One possibility is that this is just an artefact of the EFT construction and appears only in the field decomposition rather than being a true symmetry of the Lagrangian. However, this seems to be an interesting direction of further research.

Another possible way to solve the section constraint is to do a generalised Scherk-Schwarz reduction that relaxes the differential constraint to a set of algebraic relations on embedding tensor, known as quadratic constraints. For the $E_7$ covariant theory this was done in~\cite{Hohm:2014qga}. It is important to note, that as it was shown in~\cite{Dibitetto:2012rk}, the quadratic constraints are much weaker than the initial section condition, thus one may consider certain gaugings that break the section condition. These are claimed to correspond to the so called genuine non-geometric gaugings and are defined as such gaugings that do not belong to any geometric U-duality orbit. It is expected that such gaugings can be employed to stabilise moduli and construct inflationary potential~\cite{Hassler:2014mla}. Since classification of orbits becomes more and more complicated as the rank of the gauge group increases, exceptional field theories with simple duality groups can work as useful toy models for investigating common features. In this sense, the model constructed here is a nice analogue of the $E_7$ theory where one encounters pseudo-action and self-dual forms as well.

Finally, an interesting problem is to look for lifts of the known solutions of lower dimensional supergravities into EFT. Lift of the M2-brane solution into the $E_7$ supersymmetric EFT was recently found in~\cite{Berman:2014hna}. A fascinating property of the constructed lift is that the corresponding higher-dimensional solution is free of singularities.

\section*{Acknowledgments}

We would like to thank Emil Akhmedov, Andrei Marshakov, and especially Henning Samtleben for valuable discussions and useful comments. ETM would like to thank DESY and personally Jan Louis for warm hospitality during completion of part of this work. The work of IB is supported by the Russian Government program of competitive growth of Kazan Federal University and by the RFBR grant 14-02-31494.

\appendix

\section{Notations and conventions}
\label{not}

We collect here all the notations for indices used in this paper. 
\begin{equation}
\begin{aligned}
& \hat{\tt{M}},\hat{\tt{N}},\ldots =0,\ldots 10, && \mbox{11-dimensional spacetime indices};\\
& {\tt{M}},{\tt{N}},\ldots =0,\ldots 9, && \mbox{10-dimensional spacetime indices};\\
& \m,\n,\r \ldots =0,\ldots 5, && \mbox{6-dimensional spacetime indices};\\
& \ba,\bb,\bc \ldots =0,\ldots 5, && \mbox{6-dimensional spacetime flat indices};\\
& m,n,p\ldots =1,\ldots 5, && \mbox{5-dimensional internal curved indices};\\
& \un{m}, \un{n},\un{p} \ldots =1,\ldots 4, && \mbox{4-dimensional internal curved Type IIB indices};\\
& \hat{\a} =1, 2, && \mbox{SL(2) Type IIB index};\\
& M,N,K \ldots =1,\ldots 16, && \mbox{$SO(5,5)$ spinor indices labelling the extended space};\\
& i,j,k,l =1,\ldots 10, && \mbox{$SO(5,5)$ vector indices};\\
& \a,\b,\dot{\a},\dot{\b}\ldots =1,\ldots 4, && \mbox{spinor indices for each $SO(5)$};\\
& a,b,\dot{a},\dot{b}\ldots =1,\ldots 5, && \mbox{vector indices for each $SO(5)$};\\
\end{aligned}
\end{equation}

The $SO(5,5)$ gamma matrices are introduced by $16\times 16$ blocks $\g_{i\,MN}$ and $\g{}^{i\,MN}$ that satisfy the usual anticommutation relations
\begin{equation}
\g_{i\,MN}\g{}^{j\,NK}+\g{}^i{}_{MN}\g_j{}^{NK}=2\d{}^i_j \d{}^K_N.
\end{equation}
The 10-dimensional vector indices labelled by $i,j$ are raised and lowered by the $SO(5,5)$ invariant tensor $\h_{ij}$, that is basically the flat metric.

\section{Covariant field strengths}

\subsection{Gauge transformations}

The long spacetime derivative, covariant with respect to the D-bracket, was defined to be of the 
following form
\begin{equation}
\mc{D}_\m=\dt_\m-\mc{L}_{A_\m}=\dt_\m - \left[A_\m,\bullet\right]_D,
\end{equation}
where the generalised vector field $A_\m^M$ plays the role of the gauge connection. Let us now find 
how should the vector field transform in order for the derivative $\mc{D}_\m$ to be covariant:
\begin{equation}
\begin{aligned}
\left(\d_\L-\mc{L}_\L\right)\left(\mc{D}_\m V{}^M\right)&=\dt_\m\d_\L V{}^M-\mc{L}_{\d 
A_\m}V{}^M-\mc{L}_{A_\m}\d_\L V{}^M\\
&-\mc{L}_\L\left(\dt_\m V{}^M\right)+\mc{L}_\L\mc{L}_{A_\m}V{}^M\\
&=\dt_\m\mc{L}_\L V{}^M-\mc{L}_\L\left(\dt_\m V{}^M\right)-\mc{L}_{\d A_\m}V{}^M-[\mc{L}_{A_\m},\mc{L}_\L] 
V{}^M\\
&=\mc{L}_{\dt_\m\L} V{}^M-\mc{L}_{\d A_\m}V{}^M-\mc{L}_{[A_\m,\L]_E} V{}^M,
\end{aligned}
\end{equation}
where in the second line we have used the closure condition and the linearity of $\mc{L}_\L$ with 
respect to $\L$. Since the E-bracket differs from the D-bracket by a trivial transformation 
\eqref{brackets}, we may choose the transformation of $A_\m^M$ to be of the form similar to the 
conventional Yang-Mills:
\begin{equation}
\label{connectionapp}
\d_\L A_\m^M=\dt_\m \L^M-[A_\m,\L]_D{}^M=\mc{D}_\m\L^M.
\end{equation}
Since the E-bracket does not satisfy the Jacobi identity the commutator of covariant derivatives in 
general does not give a covariant expression
\begin{equation}
[\mc{D}_\m,\mc{D}_\n] =-\mc{L}_{F_{\m\n}}, \qquad F_{\m\n}{}^M 
=2\,\dt_{[\m}A_{\n]}{}^M-[A_{\m},A_{\n}]_E{}^M.
\end{equation}
We refer to the quantity $F_{\m\n}{}^M$ as a non-covariant field strength for the 1-form potential 
$A_{\m}^M$ and similar for the other potentials. Under an arbitrary variation of the gauge field $\d 
A_\m^M$ the non-covariant field strength transforms as
\begin{equation}
\begin{aligned}
\d F_{\m\n}{}^M & = 2\,\dt_{[\m}\d A_{\n]}^M-2[A_{[\m},\d A_{\n]}]_E{}^M\\
&=2\left(\dt_{[\m}\d A_{\n]}^M - [ A_{[\m},\d A_{\n]}]_D{}^M \right)+Y{}^{MN}_{KL}\dt_N(A_{[\m}^K \d 
A_{\n]}^L)\\
&=2\,\mc{D}_{[\m}\d A^M_{\n]}+Y{}^{MN}_{KL}\dt_N(A_{[\m}^K \d A_{\n]}^L).
\end{aligned}
\end{equation}
We see that if we restrict $A_\m^M$ to transform as a gauge connection~\eqref{connectionapp}, then 
the transformation of $F_{\m\n}{}^M$ contains a covariant piece and some extra terms:
\begin{equation}
\d_\L F_{\m\n}{}^M = \left(\mc L_\L F_{\m\n}\right){}^M - Y_{KL}^{MN}\, \dt_N \left( \L^K F_{\m\n}{}^L - 
A_{[\m}^K \, \mc{D}_{\n]} \L^L\right).
\end{equation}
In the spirit of tensor hierarchy the non-covariant terms here may absorbed into variation of some 
2-form $B_{\m\n}{}^{KL}$ by defining the full covariant field strength
\begin{equation}
\label{F2}
\mc{F}_{\m\n}{}^M=F_{\m\n}{}^M-Y{}^{MN}_{KL}\dt_N B_{\m\n}{}^{KL}.
\end{equation}
Its general variation takes the form
\begin{equation}
\label{trans_Fmn}
\d\mc{F}_{\m\n}{}^M=2\,\mc{D}_{[\m}\d A{}^M_{\n]}-Y{}^{MN}_{KL}\dt_N \D B_{\m\n}{}^{KL},
\end{equation}
with
\begin{equation}
\D B_{\m\n}{}^{KL}=\d B_{\m\n}{}^{KL}-\fr{1}{D(1-2\b_d)}Y{}^{KL}_{MN}A_{[\m}^{M}\d A_{\n]}^{N}
\end{equation}
(we have used the relation $Y{}^{MN}_{KL}Y{}^{KL}_{PQ}=D(1-2\b_d )Y{}^{MN}_{KL}$). It is important that 
the $B$-field transforms under $\L$-transformations in such a way that the term $Y{}^{MN}_{KL}\dt_N 
B_{\m\n}{}^{KL}$ is not covariant. Hence the expression \eqref{trans_Fmn} becomes a generalised 
tensor. Note that since the full covariant field strength $\mc{F}_{\m\n}{}^M$ differs from 
${F}_{\m\n}{}^M$ by a trivial gauge transformation, it appears in the commutator of covariant 
derivatives as well:
\begin{equation}
[\mc{D}_\m,\mc{D}_\n]=-\mc{L}_{F_{\m\n}}=-\mc{L}_{\mc{F}_{\m\n}}.
\end{equation}

Requiring that the newly introduced field strength $\mc F_{\m\n}{}^M$ transform covariantly under the 
transformations parametrized by $\L^M$ should in principle fix the transformation law $\d_\L 
B_{\m\n}{}^{KL}$. However, if we identify the field $B_{\m\n}{}^{KL}$ with the 2-form $B$-field of the 
maximal $D=5,6$ supergravities, we may expect its own gauge variation with a 1-form 
parameter $\X_\m{}^{KL}$ to modify the transformation law. The gauge variation of $A_\m^M$ would also 
be affected. Overall, we may expect the following gauge transformations of the fields corresponding 
to the ${SO}(5,5)$ and ${SL}(5)$ duality groups~\cite{Bergshoeff:2007ef}:
\begin{equation}
\label{trans_ABapp}
\begin{aligned}
\d A_\m^M &=\mc{D}_\m\L^M+Y{}^{MN}_{KL}\dt_N\X_\m{}^{KL},\\
\D B_{\m\n}{}^{KL} &= 
2\mc{D}_{[\m}\X_{\n]}{}^{KL}-\fr{1}{D(1-2\b_d)}Y{}^{KL}_{MN}\L^M\mc{F}_{\m\n}{}^N 
+3 \left(\dt_N\Y_{\m\n}{}^{N,KL}-Y{}^{KL}_{PQ}\dt_N \Y_{\m\n}{}^{P,NQ}\right).
\end{aligned}
\end{equation}
For this choice of gauge transformations, 
the covariant field strength $\mc F_{\m\n}{}^M$ transforms as a generalised vector with the 
appropriate weight $\b_d$:
\begin{equation}
\label{trans_Fmn1}
\d_\L\mc{F}_{\m\n}{}^N = \left(\mc{L}_\L\mc{F}_{\m\n}\right){}^M.
\end{equation}
Indeed, substituting the transformations~\eqref{trans_ABapp} into~\eqref{trans_Fmn} and taking into 
account the identity $Y{}^{MN}_{KL}Y{}^{KL}_{PQ}=D(1-2\b_d )Y{}^{MN}_{KL}$, one obtains $\d 
\mc{F}_{\m\n}{}^M=[\L,\mc{F}_{\m\n}]_D^M$, that is exactly~\eqref{trans_Fmn1}. 

The $\Y$ terms in the variation $\D B_{\m\n}{}^{KL}$~\eqref{trans_ABapp} were added to covariantise 
the transformation of the field strength for the 2-form field $B_{\m\n}{}^{KL}$, that we are about to 
construct. It is important, that they do not contribute to the transformation of the 2-form 
$\mc{F}_{\m\n}$. One can check that this combination of $Y$-contractions of a 
generalised tensor $\h{}^{M,KL}(=\h{}^{M,LK})$ forms a generalised tensor
\begin{equation}
\label{Psi}
\d_\L\left(\dt_N\h{}^{N,KL}-Y{}^{KL}_{PQ}\dt_N 
\h{}^{P,NQ}\right)=\mc{L}_\L\left(\dt_N\h{}^{N,KL}-Y{}^{KL}_{PQ}\dt_N \h{}^{P,NQ}\right).
\end{equation}
Together with the term $Y{}^{MN}_{KL}\dt_M \c{}^{KL}$ these appear as extended geometry analogues of 
differential forms in Riemannian geometry. Indeed, having a $p$-form $\w{}^{p}$ one does not need a 
covariant derivative to construct a $(p+1)$-form  $\w{}^{p+1}=d \w{}^{p}$. Since we have exceptional 
groups instead of $GL(D)$ one does not simply antisymmetrise the corresponding indices.

The next step is to construct such a covariant 3-form field strength for the $B$-field that its 
first term has the usual form $\mc{D}_{[\m}B_{\n\r]}{}^{KL}$. The most straightforward way to proceed 
is to start with the Bianchi identity for the covariant field strength $\mc{F}_{\m\n}{}^M$:
\begin{equation}
\label{bianchi2app}
3\mc{D}_{[\m}\mc{F}_{\n\r]}{}^M=-Y{}^{MN}_{KL}\dt_N \F_{\m\n\r}{}^{KL},
\end{equation}
where again the covariant field strength $\F$ is constructed of the non-covariant one $F$ by adding 
an extra term to be determined
\begin{equation}
\label{F3cov}
\begin{aligned}
F_{\m\n\r}{}^{KL} & = 
3\,\mc{D}_{[\m}B_{\n\r]}{}^{KL}+\fr{3}{D(1-2\b_d)}Y{}^{KL}_{PQ}\left(A_{[\m}^{(P}\dt_\n 
A^{Q)}_{\r]}-\fr13[A_{[\m},A_{\n}]_E{}^{(P}A_{\r]}^{Q)}\right),\\
\F_{\m\n\r}{}^{KL}&= F_{\m\n\r}{}^{KL}-\Phi_{\m\n\r}{}^{KL}.
\end{aligned}
\end{equation}
The reader is referred to the next section for the details of this calculation. The last term here 
will be constructed out of the next field in the tensor hierarchy, which is the 3-form 
$C_{\m\n\r}{}^{M,KL}$, with some derivatives and possible contractions with the $Y$-tensor. 

Following the analogy with the gauged supergravity we would like the transformation of the covariant 
field strength to be of the  form
\begin{equation}
\label{var_cov}
\begin{aligned}
\d \F_{\m\n\r}{}^{KL}=3\,\mc{D}_{[\m}\D B_{\n\r]}{}^{KL}+\fr{3}{D(1-2\b_d)}Y{}^{KL}_{PQ}\F_{[\m\n}{}^P\d 
A_{\r]}^Q-\D \Phi_{\m\n\r}{}^{KL}.
\end{aligned}
\end{equation}
Taking the variation of \eqref{F3cov} and transforming it to the form above we 
see, that the remaining terms can be organized into a full derivative:
\begin{equation}
\begin{aligned}
\D\Phi_{\m\n\r}{}^{KL} = &\ \d \Phi_{\m\n\r}{}^{KL}+ 3\, \dt_N \Big(-\d 
A_{[\m}^NB_{\n\r]}{}^{KL}+Y{}^{KL}_{PQ}B_{[\m\n}{}^{PN}\d A_{\r]}^{Q} \\ 
&-\fr{1}{3D(1-2\b_d)}Y{}^{KL}_{RS}\left(A_{[\m}^NA_\n^R\d A_{\r]}^S+Y{}^{RN}_{PQ}A_{[\m}^PA_\n^S\d 
A_{\r]}^Q\right)\Big).
\end{aligned}
\end{equation}
Defining the variation of the last remaining supergravity tensor field~$C_{\m\n\r}{}^{M,KL}$ to be
\begin{equation}
\label{transCapp}
\D C_{\m\n\r}{}^{N,KL} = \d C_{\m\n\r}{}^{N,KL} - \d A_{[\m}^N B_{\n\r]}{}^{KL}  -\fr{1}{3D(1-2\b_d)}
Y{}^{KL}_{RS} A_{[\m}^N A_\n^R \d A_{\r]}^S,
\end{equation}
we write 
\begin{equation}
\D\Phi_{\m\n\r}{}^{KL} = 3\, \dt_N \D C_{\m\n\r}{}^{N,KL} - 3\, Y{}^{KL}_{PQ} \dt_N \D
C_{\m\n\r}{}^{Q,PN}.
\end{equation}
This leads to the following expression for the full covariant 3-form field
strength:
\begin{equation}
\label{F3Capp}
\begin{aligned}
\F_{\m\n\r}{}^{KL} & = 3\,\mc{D}_{[\m}B_{\n\r]}{}^{KL}+
\fr{3}{D(1-2\b_d)}\,Y{}^{KL}_{PQ}\left(A_{[\m}^{(P}\dt_\n A^{Q)}_{\r]}-\fr13[A_{[\m},A_{\n}]_E{}^{(P}A_{\r]}^{Q)}\right)\\
&-3\Big(\dt_N C_{\m\n\r}{}^{N,KL} - Y{}^{KL}_{PQ}\dt_N C_{\m\n\r}{}^{Q,PN}\Big).
\end{aligned}
\end{equation}
It is straightforward to show that upon imposing the section condition the last line above does not contribute 
to the Bianchi identity~\eqref{bianchi2}. Using the equations~\eqref{trans_ABapp} and~\eqref{transCapp}, the gauge transformation of the covariant field strength can be written as  
\begin{equation}
\label{var_cov_C}
\begin{aligned}
\d \F_{\m\n\r}{}^{KL}=&\ 3\mc{D}_{[\m}\D B_{\n\r]}{}^{KL} +\fr{3}{D(1-2\b_d)}Y{}^{KL}_{PQ}\F_{[\m\n}{}^P\D 
A_{\r]}^Q\\
& -3\left(\dt_N \D C_{\m\n\r}{}^{N,KL}-Y{}^{KL}_{PQ}\,\dt_N \D C_{\m\n\r}{}^{Q,PN}\right).
\end{aligned}
\end{equation}

 Let us show explicitly that the above transformation indeed reduces to the transformation law of 
a generalised  tensor.  First fix gauge transformations for the 3-form potential to be:\footnote{Note, that in the off-shell formulation for the $SO(5,5)$ case the field strength in the 
last term here should be replaced by $\GG_{\m\n\r}{}^{KL}$.}
 \begin{equation}
 \D C_{\m\n\r}{}^{M,KL}=3\mc{D}_{[\m} 
\Y_{\n\r]}{}^{M,KL}-\mc{F}_{[\m\n}{}^N\X_{\r]}{}^{KL}+\fr{2}{3D(1-2\b_d)}Y{}^{KL}_{PQ}\L^P\mc{F}_{\m\n\r}{}^
{ QM}.
 \end{equation} Consider now the gauge transformations generated by 
 $\Y_{\m\n}{}^{N,KL}$, which give
 \begin{equation}
 \begin{aligned}
 &\d_\Y \F_{\m\n\r}{}^{KL}=\\
 &=3\mc{D}_{[\m}(\dt_N\Y_{\m\n}{}^{N,KL}-Y{}^{KL}_{PQ}\dt_N \Y_{\m\n}{}^{P,NQ}) 
-3\dt_N \mc{D}_{\m}\Y_{\n\r}{}^{N,KL}+3Y{}^{KL}_{PQ}\dt_N \mc{D}_{\m}\Y_{\n\r}{}^{Q,PN}\\
 &=-3\mc{L}_{A_{[\m}}(\dt_N\Y_{\m\n}{}^{N,KL}-Y{}^{KL}_{PQ}\dt_N \Y_{\m\n}{}^{P,NQ}) +3\dt_N 
\mc{L}_{A_{\m}}\Y_{\n\r}{}^{N,KL}-3Y{}^{KL}_{PQ}\dt_N \mc{L}_{A_{[\m}}\Y_{\n\r}{}^{Q,PN}.
 \end{aligned}
 \end{equation}
 Since equation \eqref{Psi} implies that the particular combination transforms as a generalised tensors, the above expression is identically zero.
 
 Next, we turn to the gauge transformations generated by $\X_\m{}^{MN}$, that give
\begin{equation}
\begin{aligned}
&\d_\X \F_{\m\n\r}{}^{KL}=\\
&=6\mc{D}_{[\m}\mc{D}_\n \X_{\r]}{}^{KL}+3Y{}^{KL}_{PQ}\F_{[\m\n}{}^P\dt_N \X_{\r]}{}^{NQ}+
3\dt_N(\F_{\m\n}{}^N\X_\r{}^{KL})-3Y{}^{KL}_{PQ}\dt_N (\F_{\m\n}{}^Q\X_\r{}^{PN})\\
&=6\X_{[\r}{}^{P(K}\dt_P\F_{\m\n]}{}^{L)}-6Y{}^{R(K}_{PQ}\X_{[\r}{}^{L)Q}\dt_R\F_{\m \n]}{}^P+3\dt_N \F_{[\m\n}{}^N\X_{\r]}{}^{KL}
-3Y{}^{KL}_{PQ}\X_{[\r}{}^{PR}\dt_R\F_{\m \n]}{}^Q=0,
\end{aligned}
\end{equation}
where the relation $\mc{D}_{[\m}\mc{D}_{\n]}=-\fr12\mc{L}_{\mc{F}_{\m\n}}$ and the identities \eqref{rel} were used. In addition, one should note here, that the gauge transformation parameter $\X_\m{}^{KL}$ satisfies the relation
\begin{equation}
\X_\m{}^{KL}=\fr{1}{D(1-2\b_d)}Y{}^{KL}_{MN}\X_\m{}^{MN}.
\end{equation}

Finally, one has to show that the rest indeed gives generalised Lie derivative of $\F_{\m\n\r}{}^{KL}$. The corresponding terms in the variation read
\begin{equation}
\begin{aligned}
\d_\L 
\F_{\m\n\r}{}^{KL}&=-\fr{3}{D(1-2\b_d)}Y{}^{KL}_{MN}\mc{D}_\m(\L^M\F_{\n\r}{}^N)+\fr{3}{D(1-2\b_d)}Y{}^{KL}_
{MN}\F_{\n\r}{}^M\mc{D}_\m\L^N\\
&-\fr{2}{D(1-2\b_d)}\dt_N\Big(Y{}^{KL}_{PQ}\L^P\mc{F}_{\m\n\r}{}^{QN}-Y{}^{KL}_{PQ}Y{}^{PN}_{RS}\L^R\F_{\m\n\r}{}^{SQ}\Big)\\
&=Y{}^{KL}_{PR}\L^P\dt_N\F_{\m\n\r}{}^{RN}-\fr{1}{D(1-2\b_d)}\Big(2Y{}^{KL}_{R(Q}\d{}^N_{S)}-2Y{}^{KL}_{P(Q}Y{}^{PN}_{S)R}\Big)\L^R\dt_N\F_{\m\n\r}{}^{SQ}\\
&-\fr{1}{D(1-2\b_d)}\Big(2Y{}^{KL}_{R(Q}\d{}^N_{S)}-2Y{}^{KL}_{P(Q}Y{}^{PN}_{S)R}\Big)\dt_N\L^R\mc{F}_{\m\n\r}{}^{QS}.
\end{aligned}
\end{equation}
Using the covariance condition \eqref{rel1} and the relation $Y{}^{MN}_{KL}\mc{F}_{(3)}{}^{KL}=D(1-2\b_d)\mc{F}_{(3)}{}^{MN}$ one obtains
\begin{equation}
\begin{aligned}
\d_\L \F_{\m\n\r}{}^{KL}&=\L^N\dt_N\F_{\m\n\r}{}^{KL}-\fr{1}{D(1-2\b_d)}\Big(2Y{}^{KL}_{R(Q}\d{}^N_{S)}-2Y{}^{KL}_{P(Q}Y{}^{PN}_{S)R}\Big)\dt_N\L^R\mc{F}_{\m\n\r}{}^{QS}\\
&=\L^N\dt_N\F_{\m\n\r}{}^{KL}+\fr{1}{D(1-2\b_d)}\Big(Y{}^{KL}_{SQ}\d{}^N_{R}-Y{}^{KL}_{PR}Y{}^{PN}_{SQ}\Big)\dt_N\L^R\mc{F}_{\m\n\r}{}^{QS}\\
&=\L^N\dt_N\F_{\m\n\r}{}^{KL}-\fr{2}{D(1-2\b_d)}\Big(Y{}^{N(K}_{SQ}\d{}^{L)}_{R}-Y{}^{N(K}_{PR}Y{}^{L)P}_{SQ}\Big)\dt_N\L^R\mc{F}_{\m\n\r}{}^{QS}\\
&=\mc{L}_\L\F_{\m\n\r}{}^{KL}.
\end{aligned}
\end{equation}
In the third line here we used the identity \eqref{rel} for contractions of the $Y$-tensor.

Finally, we need to check covariance of the 4-form field strength $\F_{\m\n\r\s}{}^{M,KL}$ which, however, appears in the $SL(5)$ EFT only under the following projection:
\begin{equation}
\dt_N \F_{\m\n\r\s}{}^{N,KL}-Y{}^{KL}_{PQ}\,\dt_N \F_{\m\n\r\s}{}^{Q,PN}.
\end{equation}
This is in complete analogy with the maximal gauged $D=7$ supergravity where the corresponding field appears under a particular projection by the embedding tensor.

The 4-form field strength is determined via the Bianchi identity for the covariant field strength $\F_{\m\n\r}{}^{KL}$ that reads
\begin{equation}
\begin{aligned}
4\mc{D}_{[\m}\F_{\n\r\s]}{}^{KL}=&\fr{3}{D(1-2\b_d)}Y{}^{KL}_{PQ}\F_{[\m\n}{}^P\F_{\r\s]}{}
{}^Q-3\big(\dt_N
\F_{\m\n\r\s}{}^{N,KL}-Y{}^{KL}_{PQ}\,\dt_N \F_{\m\n\r\s}{}^{Q,PN}\big).
\end{aligned}
\end{equation} 
So defined field strength for the 3-form potential $C_{\m\n\r}{}^{M,KL}$ takes the following form
\begin{equation}
\begin{aligned}
\F_{\m\n\r\s}{}^{M,KL}=&\ 4\mc{D}_{[\m}C_{\n\r\s]}{}^{M,KL}+ \left(2B_{\m\n}{}^{KL}\F_{\r\s}{}^{M}-B_{[\m\n}{}^{KL}Y{}^{MN}_{PQ}\dt_NB_{\r\s]}{}^{PQ}\right)\\
&+\fr{4}{D(1-2\b_d)} Y{}^{KL}_{PQ}\left(A_{[\m}^MA_\n^P\dt_\r A_{\s]}^Q-\fr14A_{[\m}^M[A_\n,A_\r]_E{}^PA_{\s]}^Q\right).
\end{aligned}
\end{equation}
Again, for explicit derivation of this expression the reader is referred to the next section.

\subsection{Bianchi identities}

As in the gauged supergravity the field strength for the 2-form potential $B_{\m\n}{}^{KL}$ is constructed by considering Bianchi identity for the covariant field strength $\mc{F}_{\m\n}{}^M$:
\begin{equation}
\label{bianchi2}
3\,\mc{D}_{[\m}\mc{F}_{\n\r]}{}^M=-Y{}^{MN}_{KL}\dt_N F_{\m\n\r}{}^{KL}.
\end{equation}
Let us first extract the non-covariant 3-form field strength $F_{\m\n\r}{}^{KL}$. Substituting the explicit form of $\mc{F}{}^M_{\m\n}$ we obtain for the left-hand side:
\begin{equation}
\begin{aligned}
&\mc{D}_{[\m}\mc{F}_{\n\r]}{}^M=\\
&=\mc{D}_{[\m}{F}_{\n\r]}{}^M-\mc{D}_{[\m}\left(Y{}^{MN}_{KL}\dt_NB_{\n\r]}{}^
{KL}\right)\\
&=-\dt_{[\m}\left[A_\n,A_{\r]}\right]_E{}^M-\left[A_{[\m},F_{\n\r]}\right]_E{}^M- 
\fr12Y{}^{MN}_{KL}\,\dt_N\left(A_{[\m}^KF_{\n\r]}{}^L\right)- Y{}^{MN}_{KL}\mc{D}_{[\m}\dt_NB_{\n\r]}{}^{KL}\\
&=\left[A_{[\m},\left[A_\n,A_{\r]}\right]_E\right]_E{}^M-\fr12Y{}^{MN}_{KL}\dt_N\left(A_{[\m}^KF_{\n\r]}
{}^L\right)-  Y{}^{MN}_{KL}\dt_N\mc{D}_{[\m}B_{\n\r]}{}^{KL}\\
&=-Y{}^{MN}_{KL}\dt_N\left(\mc{D}_{[\m}B_{\n\r]}{}^{KL}+A_{[\m}^K\dt_\n A^L_{\r]}- 
\fr13\left[A_{[\m},A_{\n}\right]_E{}^KA_{\r]}^L\right),
\end{aligned}
\end{equation}
where in the second line we have used the relation \eqref{brackets} between the E- and D-brackets. In the third line the relation 
\begin{equation}
Y{}^{MN}_{KL}\dt_N\mc{D}_{\m}\c{}^{KL}=Y{}^{MN}_{KL}\mc{D}_{\m}\dt_N\c{}^{KL}
\end{equation}
was used, which is valid for any symmetric generalised tensor $\c{}^{KL}(=\c{}^{LK})$. Finally, in the last line we have used the Jacobi identity for the E-bracket~\eqref{Jac_E}. Hence, we conclude that the covariant field strength for the 2-form field can be taken in the following form:
\begin{equation}
\begin{aligned}
\F_{\m\n\r}{}^{KL} & = 3\,\mc{D}_{[\m}B_{\n\r]}{}^{KL}+
\fr{3}{D(1-2\b_d)}\,Y{}^{KL}_{PQ}\left(A_{[\m}^{(P}\dt_\n A^{Q)}_{\r]}-\fr13[A_{[\m},A_{\n}]_E{}^{(P}A_{\r]}^{Q)}\right)\\
&-\left(3\,\dt_N C_{\m\n\r}{}^{N,KL} - 3\,Y{}^{KL}_{PQ}\,\dt_N C_{\m\n\r}{}^{Q,PN}\right),
\end{aligned}
\end{equation}

To construct the EFT for the U-duality group $SL(5)$ one needs a covariant field strength for the 3-form potential. The corresponding Bianchi identity takes the following form
\begin{equation}
\label{bianchi3app}
\begin{aligned}
4\mc{D}_{[\m}\F_{\n\r\s]}{}^{KL}=&\fr{3}{D(1-2\b_d)}Y{}^{KL}_{PQ}\F_{[\m\n}{}^P\F_{\r\s]}{}
{}^Q-3\big(\dt_N
\F_{\m\n\r\s}{}^{N,KL}-Y{}^{KL}_{PQ}\dt_N \F_{\m\n\r\s}{}^{Q,PN}\big).
\end{aligned}
\end{equation} 
Where the field strength for the 3-form potential $C_{\m\n\r}{}^{M,KL}$ reads
\begin{equation}
\label{F4app}
\begin{aligned}
\F_{\m\n\r\s}{}^{M,KL}=&\ 4\mc{D}_{[\m}C_{\n\r\s]}{}^{M,KL}+\left(2B_{\m\n}{}^{KL}\F_{\r\s}{}^{M}- B_{[\m\n}{}^{KL}Y{}^{MN}_{PQ}\dt_NB_{\r\s]}{}^{PQ}\right)\\
&+\fr{4}{D(1-2\b_d)} Y{}^{KL}_{PQ}\left(A_{[\m}^MA_\n^P\dt_\r A_{\s]}^Q-\fr14A_{[\m}^M[A_\n,A_\r]_E{}^PA_{\s]}^Q\right).
\end{aligned}
\end{equation}
Indeed, let us show that the LHS and RHS of the Bianchi identity match upon substituting the above expression and \eqref{F3Capp} into \eqref{bianchi3app}. Consider first the terms that depend on $B_{\m\n}{}^{KL}$:
\begin{equation}
\label{DDB}
\begin{aligned}
2\mc{D}_{\m}\mc{D}_\n B_{\r\s}{}^{KL}&=-\LL_{\F_{\m\n}}B_{\r\s}{}^{KL}\\
&=-\left(\F_{\m\n}{}^N\dt_N 
B_{\r\s}{}^{KL}-2B_{\m\n}{}^{N(K}\dt_N\F_{\r\s}{}^{L)}+2Y{}^{N(K}_{PQ}B_{\m\n}{}^{L)P}\dt_{N}\F_{\r\s}{}^Q\right)\\
&=-\left(\dt_N\left(\F_{\m\n}{}^NB_{\r\s}{}^{KL}\right)-Y{}^{KL}_{PQ}\,\dt_N\left(\F_{\m\n}{}^PB_{\r\s}{}^{QN
}\right)\right)\\
&-Y{}^{KL}_{PQ}\,\dt_N B_{\r\s}{}^{NP}\F_{\m\n}{}^Q,
\end{aligned}
\end{equation}
where we have used the $Y$-tensor identities \eqref{rel} in the third line and total antisymmetrisation of the indices $\{\m\n\r\s\}$ is understood. We see that the terms in brackets in the last line above already give precisely the $B\F$-terms in \eqref{F4app}.

Let us go further and consider the terms in brackets in \eqref{F3Capp}, that give (dropping the factor $D(1-2\b_d)$ for a while):
\begin{equation}
\begin{aligned}
&3 Y{}^{KL}_{PQ}\mc{D}_{\m}\left(A_{\n}^{(P}\dt_\r 
A{}^{Q)}_{\s}-\fr13[A_{\n},A_{\r}]_E{}^{(P}A_{\s}^{Q)}\right)=\\
&=3Y{}^{KL}_{PQ}\,\dt_{[\m} A_\n^P \dt_\r A_{\s]}^Q+Y{}^{KL}_{PQ}[A_\m,[A_\n,A_\r]_EA_\s]_D^{PQ}\\
&-3\Big([A_{[\m},Y_{PQ}A_\n^P\dt_\r A_{\s]}{}^Q]_D^{KL}+\fr23Y{}^{KL}_{PQ}[\dt_\m A_\n,A_\r]_E{}^PA_\s^Q 
+\fr13Y{}^{KL}_{PQ}[A_\n,A_\r]_E{}^P\dt_\m A_\s^Q\Big).
\end{aligned}
\end{equation}
Using the identities \eqref{rel} and \eqref{rel1}, and the Jacobi identity  
\eqref{Jac_E} the first term here and the terms in brackets can be simplified as follows
\begin{equation}
\label{DYAAA}
\begin{aligned}
&3Y{}^{KL}_{PQ}\dt_{\m} A_\n^P \left(\dt_\r 
A_{\s}^Q-[A_\r,A_\s]_E{}^Q\right)-Y{}^{KL}_{PQ}\left(\dt_N(A_\m^NA_\n^P\dt_\r 
A_\s^Q)-Y{}^{PN}_{RS}\dt_N\left(A_\m^QA_\n^R\dt_\r A_\s^S\right)\right)\\
&=\fr34Y{}^{KL}_{PQ}F_{\m\n}{}^PF_{\r\s}{}^Q-\fr34Y{}^{KL}_{PQ}[A_\m,A_\n]_E{}^P[A_\r,A_\s]_E{}^Q\\
&-Y{}^{KL}_{PQ}\left(\dt_N(A_\m^NA_\n^P\dt_\r A_\s^Q)-Y{}^{PN}_{RS}\dt_N(A_\m^QA_\n^R\dt_\m 
A_\n^S)\right)\\
&=\fr34Y{}^{KL}_{PQ}\F_{\m\n}{}^P\F_{\r\s}{}^Q+\fr32D(1-2\b_d)Y{}^{KL}_{PQ}\F_{\m\n}{}^P\dt_M B_{\r\s}{}^{QM}\\ 
&+\fr34Y{}^{KL}_{PQ}Y{}^{PM}_{RS}Y{}^{QN}_{UV}\dt_MB_{\m\n}{}^{RS}\dt_NB_{\r\s}{}^{UV}
-\fr34Y{}^{KL}_{PQ}[A_\m,A_\n]_E{}^P[A_\r,A_\s]_E{}^Q\\
&+Y{}^{KL}_{PQ} \left( \dt_N \left(A_\m^NA_\n^P\dt_\r 
A_\s^Q\right)+Y{}^{PN}_{RS}\dt_N\left(A_\m^QA_\n^R\dt_\m A_\n^S\right)\right).
\end{aligned}
\end{equation}
Here in the second line we have used the explicit expression for the non-covariant field strength 
\eqref{F2}.  Restoring the factor $D(1-2\b_d)$ we see that the first term in the last equation 
above exactly reproduces the $\F\F$ term in the Bianchi identities \eqref{bianchi3app} and the 
second term above precisely cancels the last term in \eqref{DDB}.

Now, to identify the $\dt B\dt B$-terms in $\F_{\m\n\r\s}{}^{M,KL}$ we substitute the 
corresponding contribution from \eqref{F4app} into the RHS of Bianchi identities 
\eqref{bianchi3app}. This gives
\begin{equation}
\begin{aligned}
&-3\,\dt_M\big(Y{}^{MN}_{PQ}B_{\m\n}{}^{KL}\dt_N 
B{}^{PQ}_{\r\s}\big)+3Y{}^{KL}_{PQ}\big(Y{}^{QR}_{UV}B_{\m\n}{}^{PN}\dt_N
 B{}^{UV}_{\r\s}\big)=\\
 &=3D(1-2\b_d)Y{}^{KL}_{PQ}\dt_MB_{\m\n}{}^{MP}\dt_NB_{\r\s}{}^{NQ}-3B_{\m\n}{}^{PN}Y{}^{KL}_{PQ}Y{}^{QR}_{UV}\,
\dt_{NR}B_{\r\s}{}^{UV}.
\end{aligned}
\end{equation}
The first term above is exactly what we had in \eqref{DYAAA} while the second term vanishes upon the section condition. Indeed, consider only the $Y$-tensors contracted with the double derivative
\begin{equation}
\begin{aligned}
&Y{}^{KL}_{PQ}Y{}^{NP}_{ST}Y{}^{QR}_{UV}\dt_{NR}=\big(-2Y{}^{KL}_{P(S}Y{}^{NP}_{T)Q}+2Y{}^{KL}_{Q(S}\d{}^N_{T)}+Y{}^{KL}_{ST}\d{}^N_Q\big)Y{}^{QR}_{UV}\dt_{NR}\\
&=-2Y{}^{KL}_{P(S}Y{}^{NR}_{T)Q}Y{}^{QP}_{UV}\dt_{NR}+2Y{}^{QR}_{UV}Y{}^{KL}_{Q(S}\dt_{T)R}=0,
\end{aligned}
\end{equation}
where in the first line we used the identity \eqref{rel} with respect to the indices $\{QST\}$ while in the last line the $Y$-invariance identity from \eqref{rel} was used with respect to the indices $\{NRP\}$.

Finally, using the same identities for the $Y$-tensor the remaining $AAAA$ terms can be shown to exactly match the RHS of Bianchi identities.

\bibliographystyle{JHEP}
\bibliography{bib}
\end{document}